\documentclass[sigconf,9pt]{acmart}

\usepackage[english]{babel}
\usepackage{blindtext}
\usepackage{enumitem}
\usepackage{makecell}
\usepackage[noend,ruled]{algorithm2e}
\usepackage{mathtools}

\usepackage{amsthm}
\usepackage{graphicx,subcaption}
\usepackage{caption} 

\acmYear{2023}\copyrightyear{2023}
\setcopyright{rightsretained}
\acmConference[ACM SIGCOMM '23]{ACM SIGCOMM 2023 Conference}{September 10--14, 2023}{New York, NY, USA}
\acmBooktitle{ACM SIGCOMM 2023 Conference (ACM SIGCOMM '23), September 10--14, 2023, New York, NY, USA}
\acmPrice{}
\acmDOI{10.1145/3603269.3604851}
\acmISBN{979-8-4007-0236-5/23/09}

\newcommand{\red}[1]{\textcolor{red}{#1}}
\begin{document}

\title{Underwater Acoustic Ranging Between Smartphones}

\title{{Bringing underwater  positioning  to    smart devices}}

\title{{Underwater  3D positioning  on    smart devices}}

\colorlet{red}{black}


\author{Tuochao Chen, Justin Chan and Shyamnath Gollakota}
\affiliation{%
    \institution{Paul G. Allen School of Computer Science \& Engineering}
  \institution{University of Washington, Seattle, WA, USA}
  \country{}
}

\renewcommand{\shortauthors}{Chen, Chan and Gollakota}


\email{underwaterGPS@cs.washington.edu}
\newcommand{\xref}[1]{\S\ref{#1}}
\newcommand{\squishlist}{\begin{itemize}[itemsep=1pt,parsep=2pt,topsep=3pt,partopsep=0pt,leftmargin=0em, itemindent=1em,labelwidth=1em,labelsep=0.5em]}
\newcommand{\squishend}{\end{itemize}}
\newcommand*{\tuochao}{\textcolor{red}}
\newcommand{\squishenum}{\begin{enumerate}[itemsep=1pt,parsep=2pt,topsep=3pt,partopsep=0pt,leftmargin=0em,listparindent=1.5em,labelwidth=1em,labelsep=0.5em]}
\newcommand{\squishsubenum}{\begin{enumerate}[itemsep=1pt,parsep=2pt,topsep=0pt,partopsep=0pt,leftmargin=0em,listparindent=1.5em,labelwidth=1em,labelsep=0.5em]}
\newcommand{\squishenumend}{\end{enumerate}}
\SetKw{Continue}{Continue}


\begin{abstract}
 \textcolor{red}{The emergence of water-proof mobile and wearable devices (e.g., Garmin Descent and Apple Watch Ultra)  designed  for   underwater activities like professional scuba diving, opens up opportunities for  underwater networking and localization capabilities on these devices.} Here, we present the first   underwater acoustic   positioning system for smart devices. Unlike  conventional  systems that  use floating buoys as  anchors   at known locations, we design a system where a dive leader can compute the relative   positions of all other divers, without any external infrastructure. Our  intuition is that in a well-connected network of devices, if we  compute the pairwise distances, we can    determine the shape of the network topology. By incorporating orientation information about a single diver who is in the visual range of the leader device, we can then estimate the positions of all the remaining divers,  even if they are not  within sight.  \textcolor{red}{ We address various practical problems including detecting erroneous distance estimates, addressing rotational and flipping ambiguities as well as designing a distributed timestamp protocol that scales linearly with the number of devices. }  Our    evaluations show that our distributed system running on  underwater  deployments  of 4-5 commodity smart devices   can perform pairwise  ranging and localization with  median errors of 0.5-0.9~m and 0.9-1.6~m. 

\begin{center}Project page with  code: 
{\textcolor{blue}{{{\url{https://underwatergps.cs.washington.edu/}}}}}\end{center}
    
\end{abstract}



\maketitle

\section{Introduction}

In recent years, both industry and academia have shown interest in developing underwater capabilities for mobile and wearable devices. While the research community has been investigating the potential of smart devices for communicating  underwater~\cite{sigcomm22}, tech companies~\cite{waterproof1,waterproof2,appleultra,appleultra2}  are rolling out waterproof versions made for underwater use like the Garmin Descent Watch series~\cite{garmin}. In 2022, among the noteworthy mobile releases was the {\it Apple Watch Ultra},  designed for professional scuba  diving at depths of 40 meters, complete with all the dive computer functions scuba divers need. With a depth gauge sensor, the watch provides a dive profile, water temperature readings, and even safety alerts such as decompression limits, fast ascents, and mandatory safety stops~\cite{appleultra,appleultra1,appleultra2}. 

\begin{figure}[t!]
\vskip -0.1in
    \includegraphics[width=.38\textwidth]{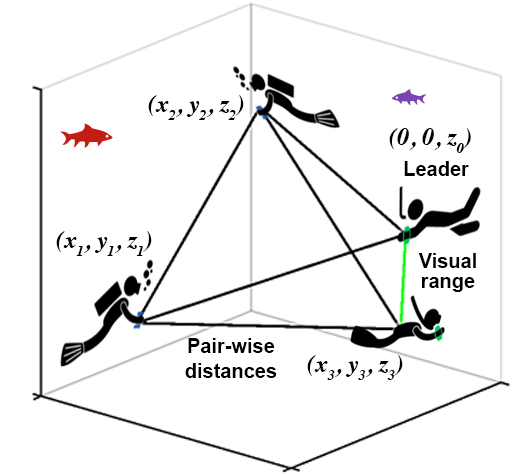}
    \vskip -0.2in
    \caption{Underwater 3D positioning for smart devices. {Our system computes the pair-wise distances between  divers using a distributed protocol, and only requires a single diver to be within the leader's visible range. Sound  can propagate  farther underwater than light.}}
 \vskip -0.15in
\label{fig:fig1}
\end{figure}

Our paper takes an  important next step in this domain   --- we  introduce the first acoustic system that enables   underwater 3D positioning capabilities on smart devices. Just as, in-air positioning systems (e.g., GPS) have been transformative for mobile devices, localization can   bring a vital new capability to underwater scenarios.  For example, maintaining close proximity with a dive  leader is critical to ensure that the  divers can help each other in the event of an emergency such as  injury or being trapped by ropes or nets~\cite{failure,net}. This can be challenging in low visibility situations such as turbid waters or during a silt out, which can cause some  divers to become  visually separated from their dive  leader~\cite{buddy,muddy}. 

Ideally, a dive instructor should be able to locate all their divers, even if they are not all within sight {(Fig.~\ref{fig:fig1})}. Because sound travels much farther underwater, current localization techniques for underwater sensor networks and robots~\cite{tracking1, tracking2,tracking3} rely on acoustic anchor nodes like floating buoys placed at known locations~\cite{vickery1998acoustic, cario2021accurate, ullah2019efficient}.  These methods involve having multiple anchor nodes that use acoustic signals to either  trilaterate the location of an underwater device or use  microphone arrays to compute  angle-of-arrival (AoA).

Deploying a dedicated  underwater anchor infrastructure  is challenging for two key   reasons: 1)  custom anchors\footnote{By  anchors,  we mean devices whose   absolute positions are apriori  known.} with floating buoys must be manually deployed before any underwater activities, which can be cumbersome to power and maintain in dynamic aquatic environments, and   2) all the divers must have a clear unobstructed path to the anchors for accurate acoustic ranging. 

We present a novel system that enables underwater localization capabilities on  mobile devices without the need for any infrastructure support (e.g., anchor buoys).  In our design, the dive leader  device  computes the 3D positions of all the other  divers  in their group   (see Fig.~\ref{fig:fig1}). Achieving this is challenging  since  localization techniques developed in our community for radios (e.g., Wi-Fi)~\cite{wifiloc2} do not translate to mobile devices in underwater scenarios. These methods either use multiple routers as  anchors at fixed locations~\cite{wifiloc7} or estimate angle of arrival  using multiple antennas at each router~\cite{wifiloc6}. While  phones and watches have multiple microphones, their separation is an order of magnitude smaller than the required half  wavelength.

Instead, we take a distributed approach to this problem. Our system assumes that there is a visible diver in range of the dive leader and that  the dive leader first orients themselves towards the visible diver. The  devices then run a distributed   protocol to find the pair-wise distances between divers. We leverage an important result in  graph embeddings~\cite{embedding1,embedding2,arvind}:  even in a  network where the number of links is linear in the number of nodes (versus quadratic in a fully-connected network),  the pair-wise distances can uniquely determine the shape of the network topology, as long as the links are well-distributed across the nodes (see \xref{sec:loss}). Combining this with the depth sensor data, allows us to calculate  the  relative diver positions in the 3D space. 

\begin{figure}[t!]
    {\includegraphics[width=.4\textwidth]{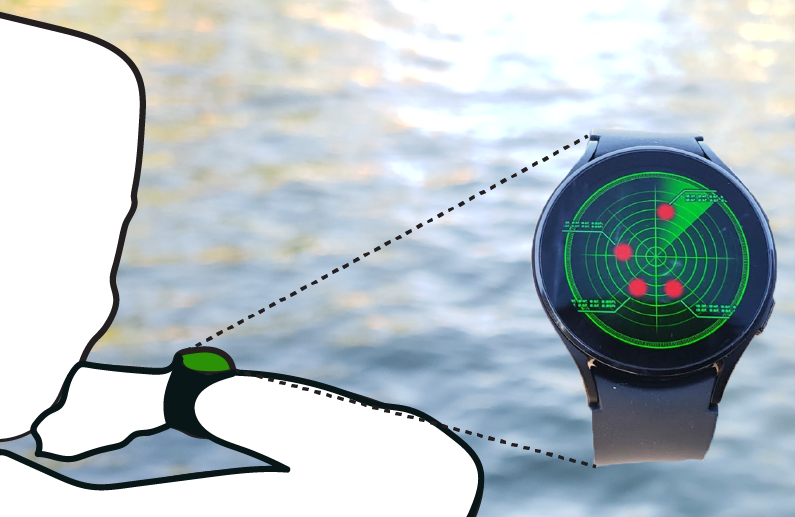}}
    \vskip -0.15in
    \caption{Conceptual  interface for our underwater system.}
 \vskip -0.15in
\label{fig:concept}
\end{figure}
    
\begin{figure}[t!]
    \includegraphics[width=.48\textwidth]{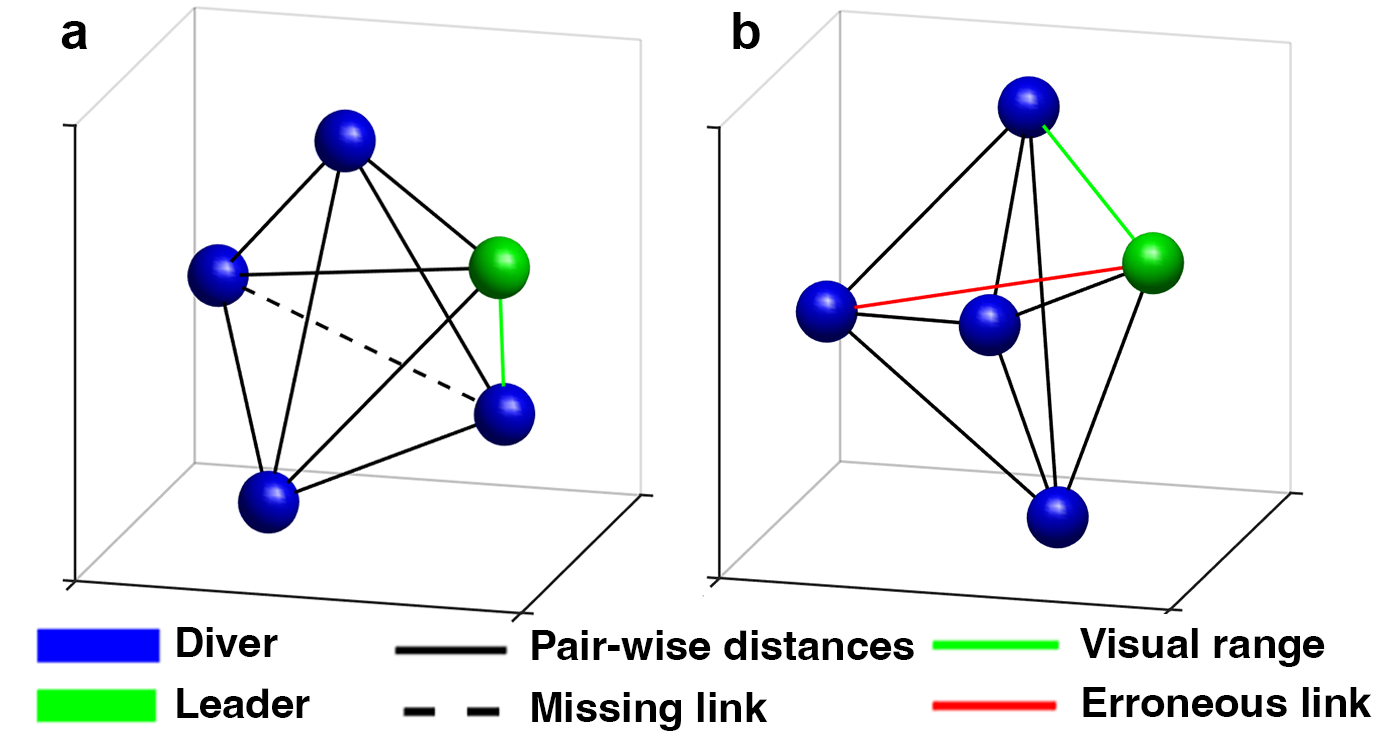}
    \vskip -0.15in
    \caption{Device localization using a network topology-based approach.   {Our system can localize (a) even with missing links in the network  and (b)  some links have severe multipath resulting in wrong distance estimates.}}
 \vskip -0.15in
\label{fig:fig2}
\end{figure}

A network topology-based  approach to underwater positioning is attractive for two reasons: 1) it does not require all the devices to be in range of each other and can work with some missing links in the network  (Fig.~\ref{fig:fig2}a), and 2) it has the ability to handle some erroneous distance measurements between devices caused by severe multipath (Fig.~\ref{fig:fig2}b).

Our design has four key components.

\squishlist

\item {\bf Pairwise distance estimation.} Underwater  channels are challenging due to multipath and noise from signal reflections between the waterbed and surface, as well as from aquatic creatures and plants, and from scattered particles in the water. The speed of sound also spreads these reflections across time causing a large delay spread~\cite{backscatterlocalization}. Mobile devices  have a limited  3-4~kHz  underwater bandwidth and  a low sampling rate of 44.1~kHz compared to commodity hydrophones~\cite{sigcomm22}.   To improve distance estimation, we use multiple microphones, such as the  2-3 microphones on smartphones or the  three-microphone array on the Apple Watch Ultra~\cite{appleultra}.  Since the time difference of arrival for the direct path between the  microphones is upper bounded by the physical   distance between them, the sample offset between the direct path in the two microphone channels  should be lower than the acoustic propagation time between them. Additionally, each microphone may have a different hardware-based noise profile.  Thus, our approach  identifies the   direct path as the earliest non-negligible peaks  {\it across} microphones whose sample offset satisfies the physical distance constraint between the   microphones on the device (see~\xref{sec:dual}).

\item {\bf Underwater topology estimation \textcolor{red}{with outlier detection.}} Given the pairwise distances, we need to  estimate the diver positions. This is challenging for two reasons. First, we might not have a fully-connected network and as a result we only have a subset of pairwise distances. Second,  some of the  pairwise distance estimates might be erroneous and outliers due to severe underwater multipath, which might significantly change the estimated network  topology.  \textcolor{red}{To address these challenges, we start with   multidimensional scaling algorithms~\cite{borg2005modern}   to estimate the topology. However even a small number of erroneous pair-wise distance estimates can significantly change the network topology output by these algorithms. To address this, we design an iterative outlier detection algorithm that drops different subsets of links and re-runs the multidimensional scaling algorithm to identify the outlier measurements and compute a uniquely realizable network topology (\xref{sec:outmeasure}).}

\item\textcolor{red}{{\bf Resolving rotational and flipping ambiguities.} While the above method can output the network topology, we still  need to address rotational and flipping ambiguities. Specifically, the whole network can rotate along the vertical axis going through the leader device, while still maintaining the depth measurements at each device. As a result, we have an infinite number of possibilities for locations for the devices in the network.    We show in~\xref{sec:flipping} that by imposing the condition that the dive leader first points towards  a  visible diver, we can  {resolve} rotational ambiguities.  This leaves us with flipping ambiguity where we have to pick between two networks that are mirror images across the line joining the leader and the device that the leader is pointing towards (Fig.~\ref{fig:ambiguity}). While the multiple microphones on smart devices do not have good AoA resolution underwater, we design a novel  technique that uses them to formulate a  binary classification problem  that addresses flipping ambiguity and uniquely determine  3D positions.} 

\item {\bf Distributed timestamp protocol.}  The challenge is that   the protocol should scale linearly with the number of devices, even when  not all devices are in range of each other and the network topology is  unknown.   At a  high level, in our protocol, the leader device  broadcasts an acoustic query  to initiate the protocol. All  other devices broadcast back a response after a fixed delay in their allocated time slots. The time slots are determined by the device ID and set with respect to when the transmission from the leader is initiated. Each device records the timestamp at which it received  messages from other devices in its range. This data is then sent back to the leader device, which uses the timestamps to compute the pair-wise distances between devices.  Further our protocol can achieve global synchronization and correct for clock drifts even when not all devices are in range of each other. 
\squishend

We implemented  our software system  and tested it in different network topologies using old Android smartphones placed in  water-proof cases, which we use for cost-effective proof-of-concept demonstration. Since     
 Android phones do not have an underwater depth sensor, we estimate depth using the pressure sensors common on smartphones (see~\xref{sec:depth}).   We evaluated our system in four different underwater environments.  
Our key  findings are as follows:
\squishlist
\item  {The median errors for pairwise 1D distance estimation  were 0.48, 0.80 and 0.86~m at 10, 20 and 35 m respectively. The median 2D localization errors were 0.8~m and 0.9~m for a 4- and 5-device network deployment, with   a protocol  latency of 1.56~s and 1.88~s. }
\item In the presence of device mobility  at 15-56~cm/s, the median   localization error for the mobile device was 0.8~m. Further, the flipping disambiguation algorithm could detect the correct positions with 90.1\% and 100\% accuracy, using signals from 1 and 3 devices with unknown positions.

\item  With missing links, the system  achieved a median  localization error of 1.0~m. The median error with occlusions resulting in  outlier distance estimates was 1.4~m.
\squishend

\vskip 0.03in\noindent{\bf Contributions.} \textcolor{red}{ Our key contributions are in the distributed system design  and the  methods to   address practical problems like outlier detection, rotational and flipping ambiguities as well as a low-latency distributed timestamp protocol.  Together this enables us to build the first underwater acoustic positioning system for smart devices.   Our  software system  does not require any  anchor infrastructure.  We evaluate our system in various underwater conditions and demonstrate its practical feasibility.  We believe that our proof-of-concept system brings 3D localization capabilities  on smart devices to the next frontier, i.e., underwater settings.}

\section{System Design}
At a high level, {our distributed 3D localization system  measures the pairwise distances.} 
\textcolor{red}{Since   RF signals attenuate quickly underwater~\cite{mary2021systematic} and light is susceptible to turbid water and high-ambient light~\cite{lacovara2008high}, we use  acoustics.}
The distance $d$ between two devices is, $c\Delta t$, where $c$ is the speed of sound and $\Delta t$ is the time-of-flight.  The  underwater sound speed  can be approximated using  Wilson's  equation~\cite{wilson1960equation}: $c = 1449 + 4.6T - 0.055T^2 + 0.0003T^3 + 1.39(S - 35) + 0.017D$. Here, $T, S, D$ are the temperature,  salinity, and  depth. The depth limit for {recreational} divers is {40}~m~\cite{scubadepth,scubadepth2}. Prior work~\cite{kuperman2007underwater} shows that at these depths, the maximum change in the speed of sound  is $30 m/s$, which is  only a $2\%$ relative error at $1500 m/s$. One  can improve  accuracy by using  depth sensors and   configuring the temperature and salinity   for different water bodies. 


If we know the timestamps at which the sender and receiver sent and received the acoustic signals, we can compute $\Delta t$. Our   system has three key components: a topology-based localization technique, a method to estimate  pairwise distances for challenging underwater multipath channels and a distributed timestamp protocol that operates in unknown network topologies. 

\subsection{Topology-based localization}\label{sec:localization}


At a high level, given a large number of  nodes in the 3D space and the exact pairwise distances between all the nodes, one can uniquely determine the shape and network topology~\cite{embedding1,embedding2}. There are four key challenges in using this  underwater.

\squishlist
\item {\it Pairwise distance estimates.} Instead of outputting the extra pairwise distances, our system  has   median errors of 0.5-0.9~m.
\item {\it Occlusions.} { Links blocked by rocks, marine life or humans, or with severe multipath lead to large distance estimate errors. }
\item {\it Missing links.} Some divers may be out of range of each other.
\item {\it Ambiguity.} In addition to  network shape, for 3D positions, we need to address rotation and flipping ambiguities.  
\squishend



Say, there are $N$ devices, including the leader and we have the  matrix $D \in \mathcal{R}^{N\times N}$, where  $D_{i,j}$ represents the pairwise distances between  device $i$ and $j$. The depth of each device is also known from onboard sensors as $h_i$. Our objective is to estimate 3D positions, $P_i = [x_i, y_i, z_i], \forall i \in [0, N-1]$,  to minimize a  stress function $S(\cdot)$.
\begin{align*}
&\min_{P_i}\quad S(P_0, \dots, P_{N-1}) = \sum_{0\leq i< j < N} w_{i,j} (D_{i,j} - \|P_i - P_j\|)^2
\end{align*}
Where $z_i = h_i$ and $P_0$ is the leader position. $w_{i,j}$ is the element of a symmetric and non-negative weight matrix. When the link between $i$ and $j$  exists, $w_{i,j}$ is 1 but is 0 for missing links. We break the problem down into three separate steps: (1) project onto the 2D space, (2) estimate the topology in that 2D space, and (3) resolve any ambiguity regarding rotation and flipping.



\subsubsection{Projection to 2D space}
Given that we get the depth  measurements from the on-device sensors,  the 3D localization problem can be simplified to a 2D localization problem using  projection~\cite{projection}. We can compute the pairwise distances  between devices $i$ and $j$ in the projected 2D space as, $D^{2D}_{i,j} = \sqrt{D_{i, j}^2 - (h_i - h_j)^2}$. 
After projection, the problem now is to optimize the 2D locations,  $P^{2D}_i = [x_i, y_i]$: 
$$\min_{P^{2D}_i}\quad  S(P^{2D}_0, \dots, P^{2D}_{N-1}) = \sum_{0\leq i< j < N} w_{i,j} (D^{2D}_{i,j} - \|P^{2D}_i - P^{2D}_j\|)^2$$

\subsubsection{Handling link loss}\label{sec:loss}
Link losses can be caused by being out of range, packet drops, and  blocking.  When a link loss happens, the pairwise distance measurement for  this link is  missing. In this setting, we  set the weights $w_{i,j}$ to 0 for missing links. We then solve the above  problem as a  multi-dimensional scaling optimization~\cite{borg2005modern}. Specifically, we use the  weighted SMACOF (Scaling by MAjorizing a COmplicated Function) algorithm~\cite{de2009multidimensional, koledoye2017mds}. Different from the steepest descent approach which directly applies iterative minimization on the stress function,  SMACOF finds a  convex function $\mathcal{T}(\cdot)$ that  majorizes the  stress function $S(\cdot)$, i.e., $\mathcal{T}(\cdot) \ge S(\cdot)$ ~\cite{de2009multidimensional} and iteratively minimizes $\mathcal{T}(\cdot)$ at each step. Compared to other approaches~\cite{de2005applications}, SMACOF  performs  better  in terms of accuracy and rate of convergence.

\begin{figure}
    \centering
    \includegraphics[width=.35\textwidth]{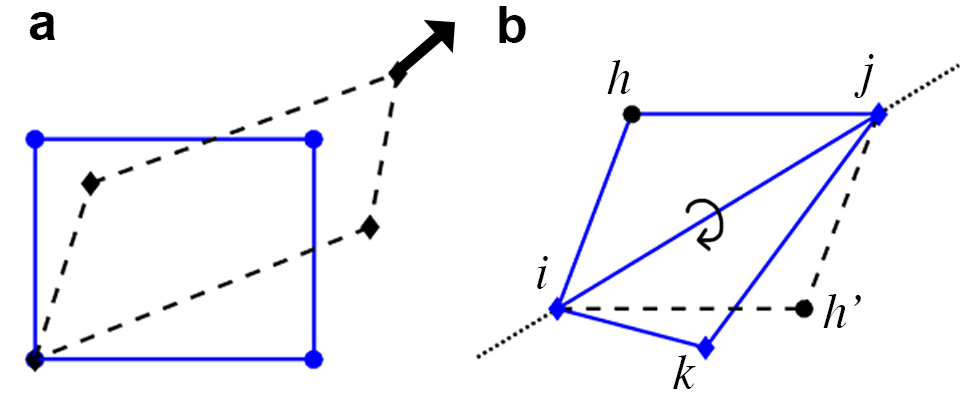}
    \vskip -0.15in
    \caption{(a) Non-rigid graph. (b) Partial reflection.}
    \vskip -0.15in
    \label{fig:theory}
\end{figure}

\vskip 0.05in\noindent{\bf Requirements.} The  topology of  a fully connected network with $N$ nodes can be uniquely determined since the ${n \choose 2} $ links constrain its shape. However, we do not need a quadratic number of links to  determine the  shape. At a high level, $N$ nodes in a 2D space have $2N$ degrees of freedom. Since the 3 degrees of freedom from rotation and translation cannot be determined by  pairwise distances, we have $2N-3$  degrees of freedom~\cite{arvind}, which is linear in the number of nodes.  In fact, a graph is defined as rigid if it does not admit a continuous deformation other than global rotation, translation, and reflection. Fig.~\ref{fig:theory}(a) shows an example graph that is not rigid and can be continuous deformed.  Laman's theorem~\cite{laman1970graphs} provides the necessary and sufficient conditions: 
{\it A graph with $n$ nodes and $2n-3$ links is rigid in two dimensions if and only if no subgraph with $n^{\prime}$ nodes has more than $2n^{\prime} - 3$ links.}

However, rigid graphs are still susceptible to discontinuous non-uniqueness, such as partial reflections. Fig.~\ref{fig:theory}(b) shows an example where one node has two possible configurations corresponding to a reflection across a set of mirror nodes. For $N=3$, three links that satisfy the triangle inequality, uniquely determine a triangle. For $N>3$ nodes, localizability is used in graph theory to describe if the graph can be  uniquely realized given the distance information. \cite{arvind} provides the necessary and sufficient condition:
{\it A graph is uniquely realizable in two dimensions iff  it is redundantly rigid and   deleting any two nodes results in a connected graph. }

A graph is redundantly rigid if it can remain rigid upon removal of any single link. If the graph satisfies the above condition, then it can be uniquely determined. In practice, however, the pairwise distances are estimated and not exact. As a result, we have errors in the 2D localization results. In~\xref{sec:sim}, we simulate different network topologies and present analytical results.

\begin{algorithm}[t!]
\caption{\tuochao{Outlier Detection} }
\label{algo:alg3}
\DontPrintSemicolon
\textbf{Input:} Set of nodes ${0, 1, \dots N - 1}$ \\ Pairwise 1d ranging matrix $D = [d_{i,j}]_{M\times M}$\\
Maximum dropping outlier number $O_{max}$\\
\textbf{Output:} 2D location of the nodes $P = [p_i]_{N\times 1}$\\
$W_0 \gets Ones(M, M)$; $//$ Init the weight matrix with all ones \\ 
$E_0, P_0 \gets SMACOF(D, W_0)$; $//$ $E_0$ is normalized stress function, $P_0$ is output of 2D positions\\
\If{$E_0 < 1.5m$}{
$P \gets P_0$;\\
\Return $P$\\
}
\For{$n_{drop} \leftarrow 1$ to $O_{max}$} {
$E_{min} \gets E_0$ \\
$P_{min} \gets P_0$ \\
\For{$s_{drop} \leftarrow Subsets(M, n_{drop})$   } { 
$//$  Subsets() outputs all possible drop subsets of $M$ links with the size $n_{drop}$ \\
$W \gets W_0$; \\
$W(s_{drop}) \gets 0$; \\
$E', P' \gets SMACOF(D, W_0)$; \\
\If{$E_0 - E' > 0.9*E_0$ and $\ E' < E_{min}$ }{
$E_{min} \gets E';$ \\
$P_{min} \gets P';$ \\ 
}
}
\If{$E_{min} < 1.5m$}{
$P \gets P_{min}$;\\
\Return $P$\\
}
$E_{0} \gets E_{min}$ \\
$P_{0} \gets P_{min}$ \\
}
\Return $P_{0}$\\
\end{algorithm}

\subsubsection{Handling outlier measurements.}\label{sec:outmeasure}
Outlier  measurements can occur when the direct path is blocked or severely attenuated, and multi-path is mistaken for the direct path. Even a small number of outliers can significantly change the  topology since our optimization function gives equal weight to every pairwise distance. While~\cite{blouvshtein2018outlier} uses the triangle inequality to detect  outliers, in our deployments, the outlier errors are often not large enough to break the triangle inequality. \cite{forero2012sparsity} uses  regularization to jointly optimize the topology and outlier detection. However, this is known to be very sensitive to the selection of regularization parameters~\cite{blouvshtein2018outlier}.

Our intuition instead is that in scenarios without outliers, the pairwise distance  errors as well as the output of the stress function in the SMACOF algorithm are both within  certain ranges. So we normalize the output of the stress function  $S(\cdot)$ with the number of links in the network. If this normalized value surpasses a threshold (1.5 in our implementation), we assume that there are outliers.  To detect the outliers, we iteratively drop different subsets of  links by setting their corresponding weights $w_{i,j}$ to  0 and re-run the SMACOF algorithm. If the new normalized stress function have significantly decreasing, the dropped links may be the outliers. We repeat this process with more outlier links until the stress function does not see a significant reduction (we set this threshold  to 90\%). Since dropping too many links  may lead to the unstable output~\cite{forero2012sparsity}, we set a maximum number of outliers to 3.   Further, we cannot drop all subsets of links since dropping some subsets may make the graph not uniquely realizable.  Thus we only run the SMACOF algorithm  when the resulting graph is still uniquely realizable.

\begin{figure}
    \centering
    \includegraphics[width=.48\textwidth]{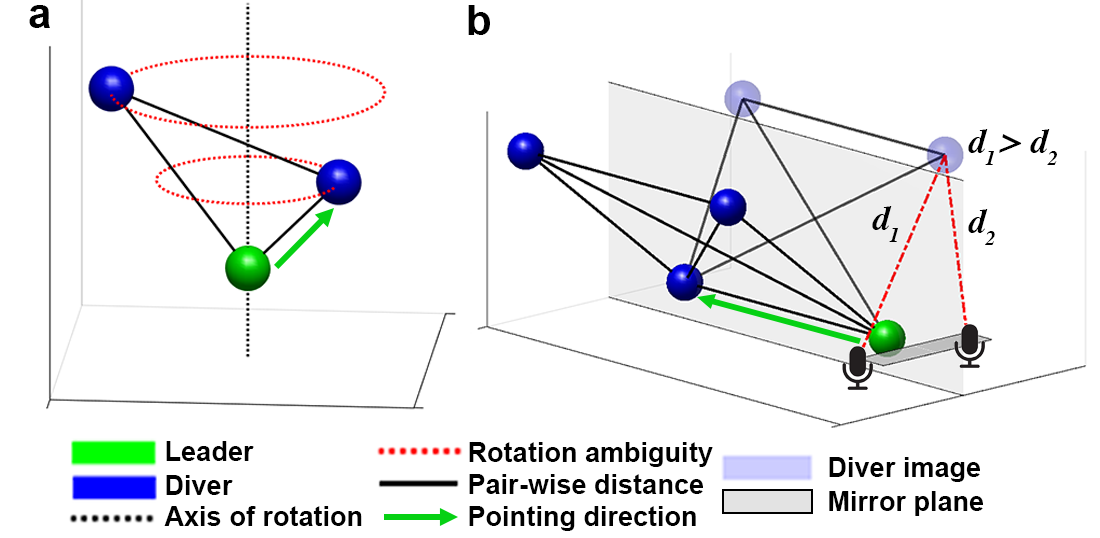}
    \vskip -0.15in
    \caption{(a) Rotational and (b) flipping ambiguity.}
    \vskip -0.15in
    \label{fig:ambiguity}
\end{figure}

\begin{figure*}[t!]
    \includegraphics[width=.24\textwidth]{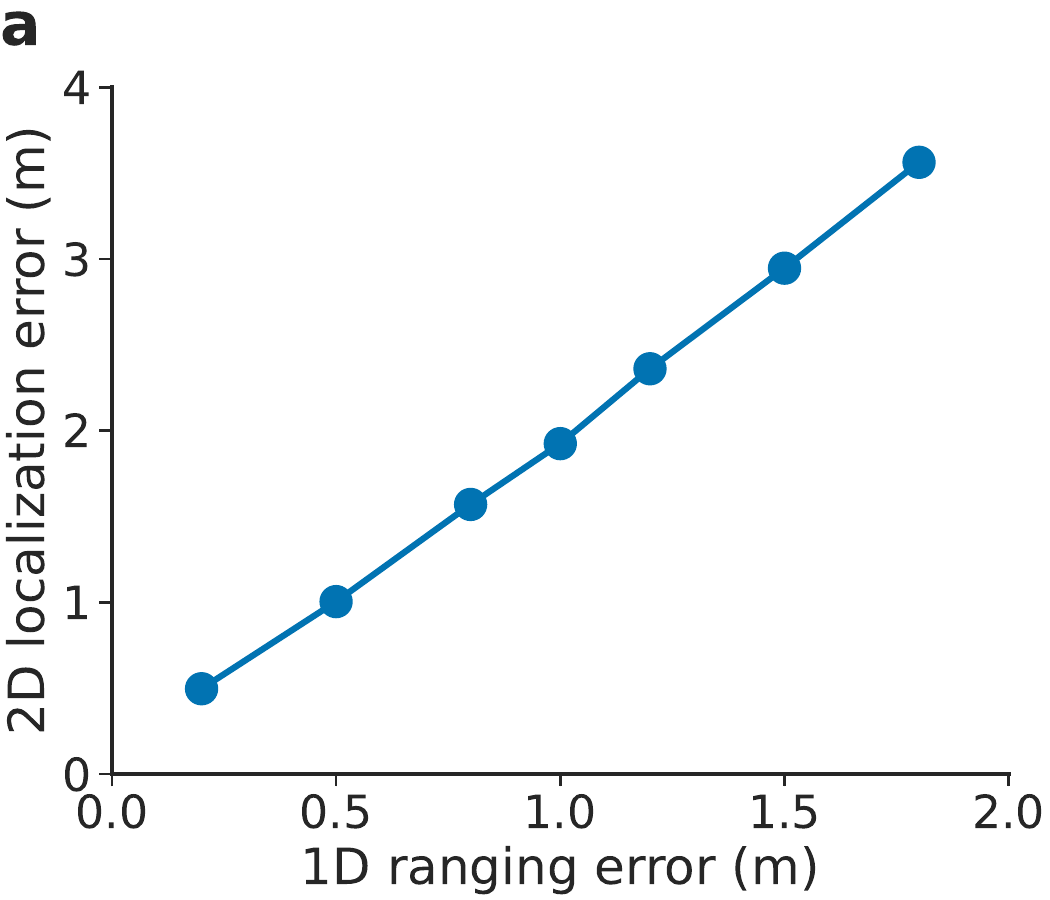}
    \includegraphics[width=.24\textwidth]{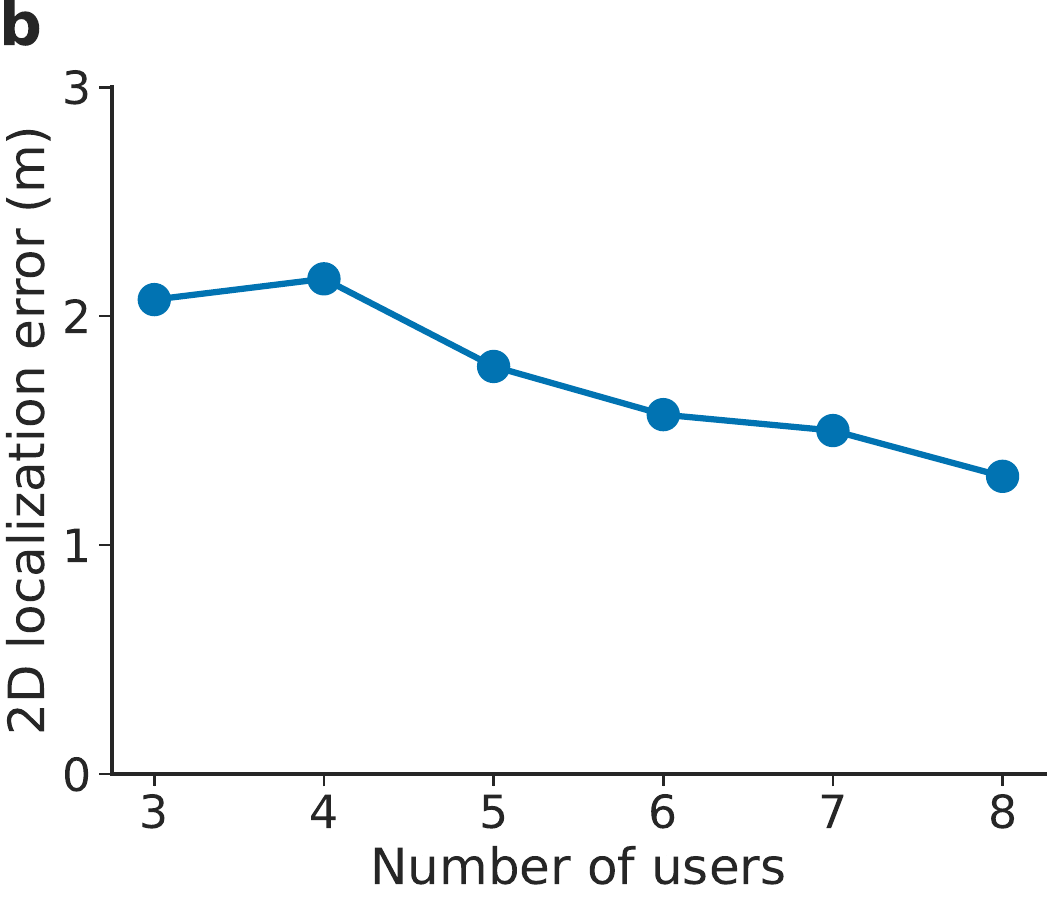}
    \includegraphics[width=.24\textwidth]{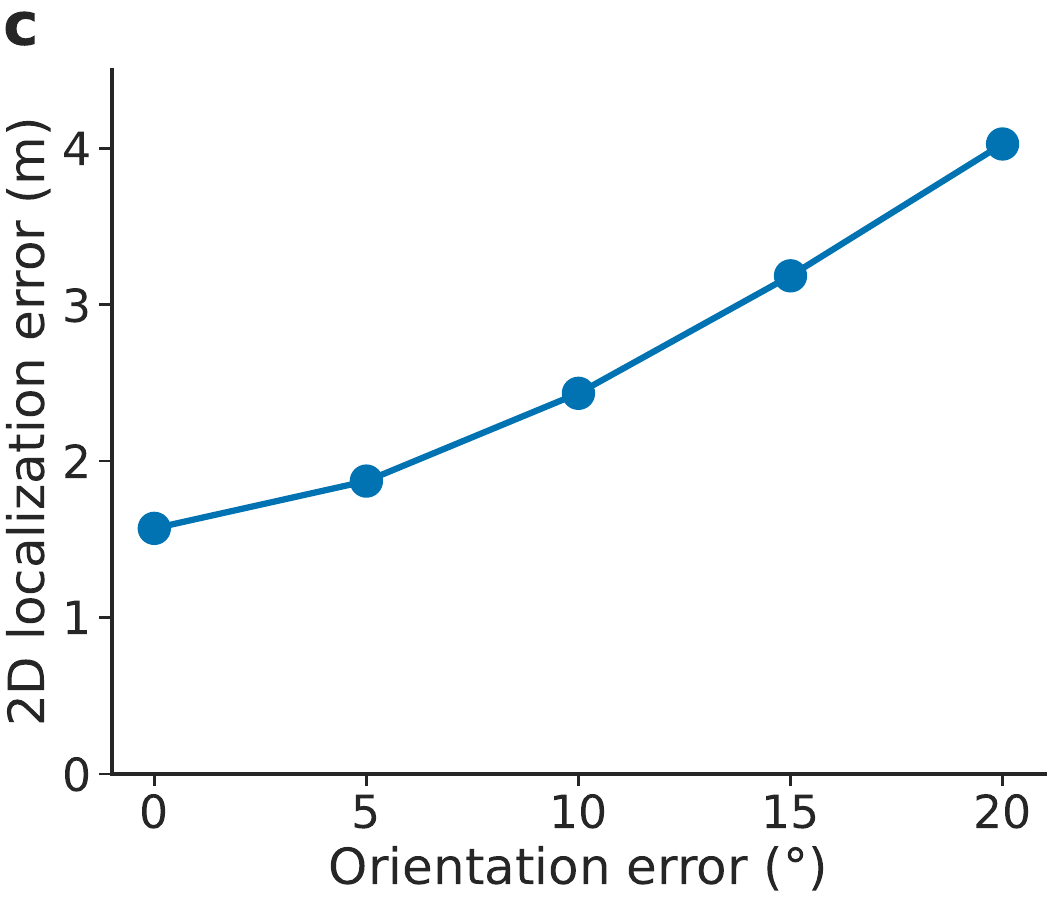}
    \includegraphics[width=.24\textwidth]{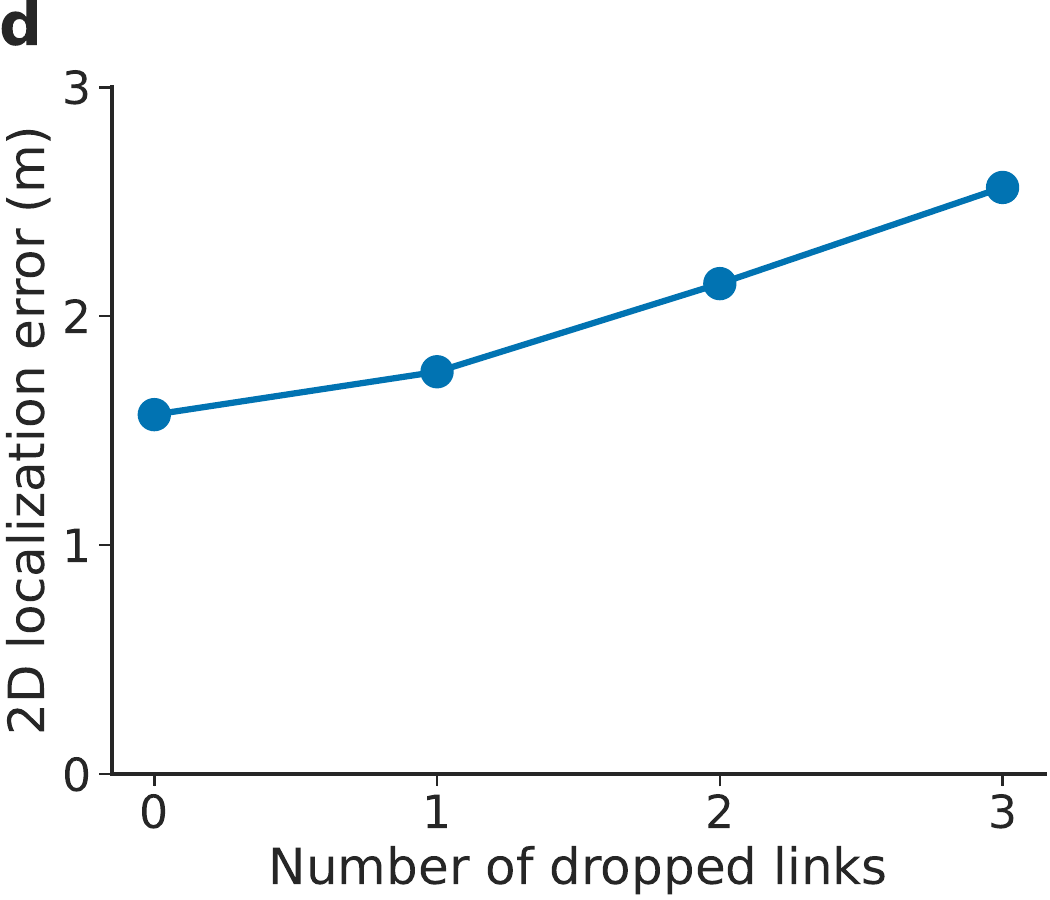}
    \vskip -0.15in
\caption{Analytical evaluation of mean error versus different parameters in a $60\times60\times10m$ 3D space.}
\vskip -0.15in
\label{fig:analysis}
\end{figure*}


\subsubsection{Rotation and Flipping ambiguity}\label{sec:flipping}
The  estimated network topology from the above optimization still has rotation and flipping ambiguities.  A rotation ambiguity occurs since as shown in Fig.~\ref{fig:ambiguity}a the topology  can rotate while still satisfying the height requirements.  This  rotational ambiguity can be addressed by rotating the edge between leader and the pointed device along the arrow in the figure, since we require the leader is oriented themselves towards the device. This leaves us with flipping ambiguity where we have to pick between two networks that are mirror images across the line joining the leader and the device that the leader is pointing towards. To resolve this,  we leverage the dual microphones in the leader's device. At a high level, while the physical separation of the two microphones on mobile devices is not large enough to provide a good AOA resolution in underwater scenarios,  we can still them to perform a simple binary classification task by determining at which microphone the diver's signal arrives first. For nodes that are to the left (right) side of the line in Fig.~\ref{fig:ambiguity}b, the left (right) microphone receives the signal before the right (left).  We can use this information to resolve flipping ambiguity.  Specifically, when the signal from diver $i$ arrives, we use the  dual-microphone channel estimation algorithm in~\xref{sec:dual}   to estimate and compare  the direct paths (we denote the direct path indexes in dual microphones are $m_i$ and $n_i$). However, in the real-world, this might be incorrect due to multipath.  To boost  flipping disambiguation accuracy, we use a voting mechanism to jointly estimate the flipping from all the diver signals (user $2, 3, \dots N-1$) (excluding the leader 0 and the pointed user 1). Since flipping gives two options:  $\{P_i\} = \{[x_i, y_i, z_i]\}$ or $\{P_i^{\prime}\} = \{[x_i^{\prime}, y_i^{\prime}, z_i^{\prime}]\}$, we compute the function, 
$$V(\{P_i\}) =   \sum_{i\in [2, N-1]} sgn(m_i\!-\!n_i) sgn\Big((x_i \!- \!x_0)(y_1\! - \!y_0) \!- \!(y_i \!- \!y_0)(x_1 \!- \!x_0)\Big).$$
If $V({\{P_i\}}) > V(\{P^{\prime}_i\})$ we output $\{P_i\}$, otherwise we output $\{P^{\prime}_i\}$. \textcolor{red}{Note that the leader only needs to point the device to the nearby diver and does not need to rotate  to a different position or angle.}

\subsubsection{Analytical evaluation}\label{sec:sim}
To understand our topology-based algorithm, we analyze it in simulation. We  randomly generate N devices in a $60\times60\times10m$ 3D space. We place the leader at the center of the 3D space and randomly generate its height. User 1 is generated with the distance between leader and user 1  randomly selected between 4 and 9~m. We then randomly generate the positions for the remaining divers in the 3D space. We add  uniformly-distributed errors to the ground-truth as our measurements: $[-\epsilon_{1d}, \epsilon_{1d}]$ for pairwise distances, $[-\epsilon_{h}, \epsilon_{h}]$ for height, and $[-\epsilon_{\theta}, \epsilon_{\theta}]$ for pointing angle. For each test, we randomly generate 200 samples and report the mean  2D localization error across all divers, excluding the leader. Fig.~\ref{fig:analysis}a plots 2D localization error as a function of the error in pairwise distances. We set $N = 6$,  $\epsilon_{h} = 0.4m$, and  $\epsilon_{\theta} = 0$. As expected,  errors in pairwise distances translate to larger 2D localization errors.  Fig.~\ref{fig:analysis}b shows that as the number of devices $N$ increases, the 2D localization error decreases. Here, we set $\epsilon_{h} = 0.4m$,  $\epsilon_{\theta} = 0$, and  $\epsilon_{1d} = 0.8m$. Fig.~\ref{fig:analysis}c sees a trend with the pointing error with the other parameters set to:  $N = 6$, $\epsilon_{h} = 0.4m$, and  $\epsilon_{1d} = 0.8m$. Fig.~\ref{fig:analysis}d shows the results with different number of link drops where  $N = 6$, $\epsilon_{h} = 0.4m$, $\epsilon_{1d} = 0.8m$, and  $\epsilon_{\theta} = 0$. 



\subsection{Pairwise distance estimation}\label{sec:dual}

{ To compute these distances, we estimate the exact timestamp when the acoustic signal arrives. This is challenging  since the direct path can be severely attenuated underwater and thus,  we can not rely on the assumption that the highest peak or the first non-negligible peak in the multipath profile is the direct path.}  Fig.~\ref{fig:dual_chan} shows that there can be some peaks before the direct path with amplitude greater than the average noise level ("Wrong peak" in Fig.~\ref{fig:dual_chan}). 

To reduce the probability of picking  these wrong peaks, we use the two microphones on the mobile devices. The basic idea of our joint synchronization algorithm  is that the time difference of arrival at the microphones (e.g., bottom and top  microphones on phones) is physically constrained by the distance between them. Thus, the sample offset between the direct path at the top microphone channel and the direct path at the bottom microphone should be lower than the acoustic propagation time between the two microphones (two black cross symbols in Fig.~\ref{fig:dual_chan}). Furthermore, the multipath created by the water-proof  case  is different at the two microphones. Finally, the two microphones may have a  different noise profile.

\begin{figure}[t!]
    \includegraphics[width=.45\textwidth]{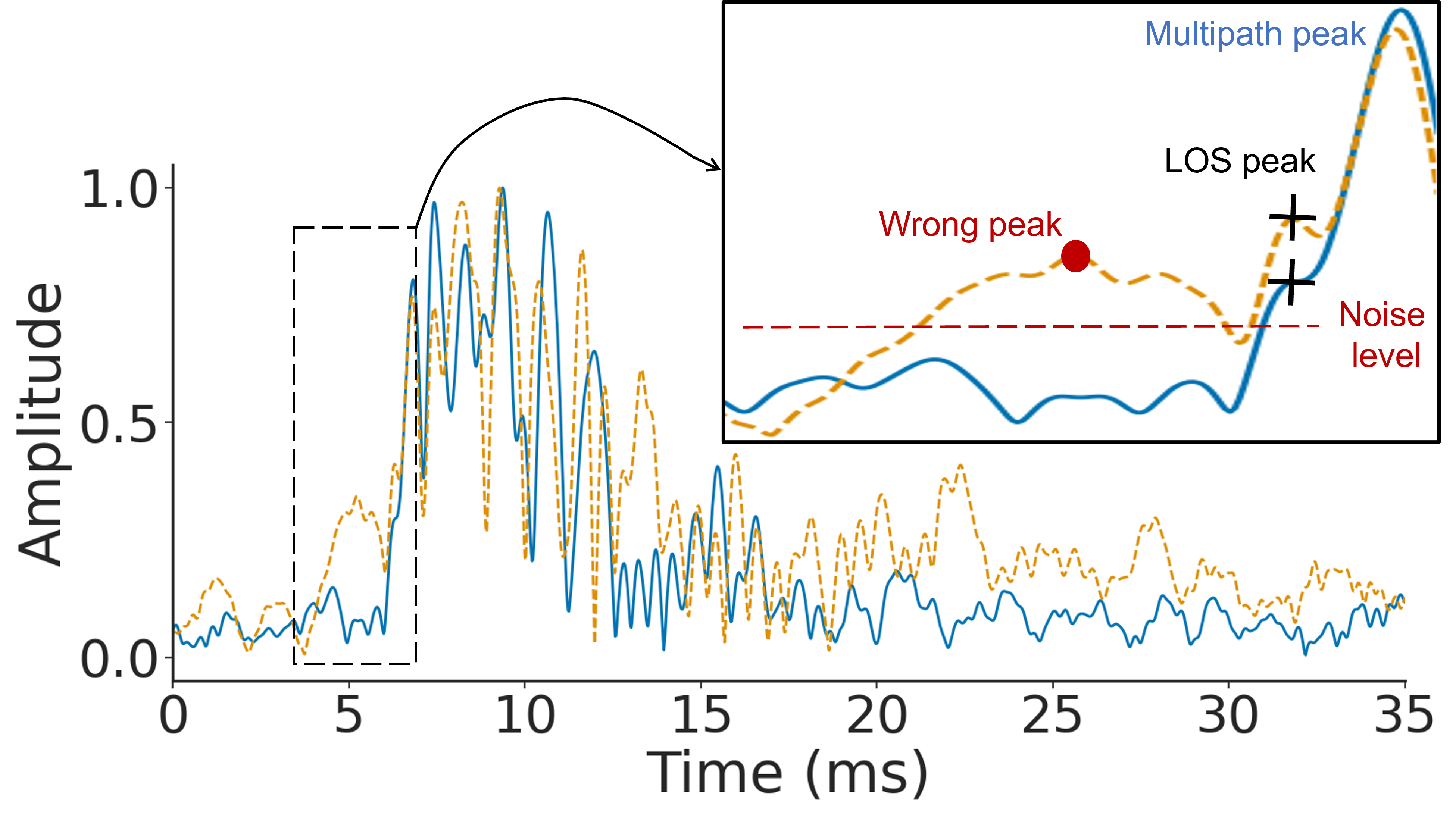}
     \vskip -0.15in
\caption{The yellow and blue curves correspond to the channel estimates at  two microphones.}
 \vskip -0.15in
\label{fig:dual_chan}
\end{figure}

Thus, our algorithm  identifies the direct path as the earliest non-negligible peaks in both channels whose sample offset satisfies the physical distance constrain  between the two  microphones. Specifically, we denote the estimated channel for the  first microphone as $h_1(n)$ and  the second microphone as $h_2(m)$, where $n$ and $m$ are the channel tap numbers. Then, we normalized $h_1$ and $h_2$ to be between  0 and 1. We then check whether the sample $n$ in channel $h_1$ is a peak. Next, we estimate the channel noise level for the two microphone channels by calculating the average power in the last 100 channel taps, respectively denoted as $w_1$ and $w_2$. 
\begin{align*}
&\min_{\tau_{LOS}}\quad \tau_{LOS} = (n+m)/2, \quad\forall m,n \in [0, L/fs)\\
& \begin{array}{r@{\quad}r@{}l@{\quad}l}
s.t.&  & h_1(n) > w_1 + \lambda,
 h_2(m) > w_2 + \lambda, \\
& & IsPeak(n, h_1) \cap IsPeak(m, h_2),  
 |n - m| \leq d/c
\end{array}
\end{align*}
 Here $\tau_{LOS}$ is the delay of the direct path, $n$ and $m$ are  the tap numbers in the channels $h_1$ and $h_2$. $w_1$ and $w_2$ are the estimated noise levels in these two channels. $\lambda$ is a conservative parameter (we set it empirically to 0.2). $d$ is the physical distance between the two microphone \tuochao{($d=16cm$ in our implementation)},  $c$ is the  speed of sound  and $L$ is the entire length of the channel (1920 samples). 


\subsubsection{ Signal processing pipeline.} 
We use OFDM symbols between 1-5~kHz as the preamble. We use this frequency band given the underwater response of mobile devices~\cite{sigcomm22}. We fill the OFDM bins with a ZC sequence~\cite{wen2006cazac} which is  phase-modulated and orthogonal to its delayed version~\cite{zhang2020endophasia}.  ZC-modulated OFDM symbols can achieve much better performance than their well-known counterpart, chirps~\cite{sesia2011lte, cai2018accurate}. We then concatenate 4 such identical OFDM symbols and multiply each with a PN sequence with different signs ([1, 1, -1, 1]), to  increase robustness to noise~\cite{nasir2010performance}. 
Between  each OFDM symbol, we insert a cyclic prefix to avoid inter-symbol interference. The length of each OFDM symbol is 1920 samples and the length of the cyclic prefix is $540$ samples. \textcolor{red}{Our preamble synchronization algorithm at the receiver is composed of three steps.}  

\tuochao{First, we perform cross-correlation between the microphone stream and the sending preamble. In the presence of a preamble, this results in a correlation peak.  However, the height of this peak could vary a lot as the SNR decreases at long distances. Meanwhile, some underwater spiky noise like bubbles would also cause high peaks in the cross-correlation, leading to plenty of false positives. }

To address this we use auto-correlation. Since our preamble has 4 OFDM symbols that are encoded with a 4-bit sequence, we split the received signal into 4 segments corresponding to the 4 OFDM symbols, multiply each segment by the 4-bit sequence, and apply correlation among them~\cite{nasir2010performance}. Auto-correlation is helpful because the spiky noise rarely has such a complex encoded pattern (PN sequence) and since the 4 received OFDM symbols suffer from nearly the same multi-path, the correlation value between two received OFDM symbols would be much higher than the correlation value between the received and transmitted symbols. We set a threshold of 0.35 in our design for valid preamble detection.

 Due to the severe underwater multi path profiles, the side-lobe height in the correlation curve is usually higher than the direct path. Hence,  coarse synchronization error based on only correlation is usually hundreds of samples, corresponding to over 6~m  error. To achieve more fine-grained synchronization, we apply channel estimation, where we leverage the channel profiles to identify the direct path.  While MUSIC-like estimators~\cite{cai2018accurate} could achieve super-resolution channel profiles, the signal space decomposition is difficult due to the extremely dense underwater channel and it has a high computational complexity. Therefore, we use an LS channel estimator~\cite{kewen2010research}. Specifically, based on the coarse synchronization of cross-correlation and auto-correlation, we segment out 4 received OFDM symbols $y_1, y_2, y_3, y_4$ from the microphone stream. Then we apply  FFTs on these 4 symbols to get $Y_1, Y_2, Y_3, Y_4$. We denote the FFT of the original OFDM symbol before multiplication with PN sequence by $X$ and denote the PN sequence by $PN_1, PN_2, PN_3, PN_4$.  The channel model can be written as $Y_i(k) = H(k)(PN_i\cdot X(k)) + N_i(k)$, where $k$ represents the $k^{th}$ frequency bin. The estimated channel is $\hat{H}(k) = \frac{1}{4}\sum_{i=1}^4 \frac{1}{PN_i}\cdot X(k)^{-1} Y_i(k)$. 



\subsubsection{Low-level audio timing} {Our distributed timestamp protocol in \xref{sec:protocol} requires each of the responding devices to transmit at a fixed time after it receives the message from the leader or other devices. 
A key challenge in achieving this is that we  need to  address the buffer delays. Specifically, at each  device, the microphone and speaker buffers  are not synced with  each other~\cite{android1}. These   buffers  are filled in   independently by the OS. Thus, we do not know the  timestamps corresponding to the  samples in the two buffers.} 
At a high level, we use  the  low-level audio timing in the OS to achieve self-synchronization between the microphone and speaker buffers on each device. { During initialization, when the microphone and speaker data streams are open, an initial calibration signal is sent from the speaker to its own microphone (green signal in Fig.~\ref{fig:pipeline1}). Then the offset between the speaker  and microphone buffers, $\Delta n$, is estimated by subtracting the microphone buffer index when  the calibration signal is detected and the speaker buffer index when the calibration signal is written (as shown in Fig.~\ref{fig:pipeline1}). Once we open the microphone and speaker data streams, we do not close them so as to keep this offset constant. When the microphone detects the leader at index m,  the device write the response message at index $n$ in speaker buffer where $n = m - \Delta n + fs \cdot t_{reply}$ ($fs$ is the sampling rate). Thus, the device can reply at a desired interval $t_{reply}$ after receiving the message at the  microphone (see Appendix). }


\subsection{Distributed timestamp protocol}\label{sec:protocol}

An approach is for each pair of devices to independently measure their pairwise distances. This scales   quadratically with the number of devices. Our protocol should  satisfy four key requirements.
\squishlist
\item {\it Efficiency.} Since divers move, it is critical for  the protocol  to compute pairwise distances across the  network in 1-2~s.
\item {\it Collisions.} {While there is no global  clock underwater, the protocol should avoid  packet collisions between devices.}
\item {\it Unknown topology.} Since  our goal is to find the  topology shape, we can not assume a known topology. Further, the protocol has to work  in networks that are not fully-connected. 
\item {\it Not all devices are  connected to leader.} The protocol should work even when not all devices can hear the leader, making time synchronization  with the leader challenging.

\squishend

\begin{figure}[t!]
    \includegraphics[width=.48\textwidth]{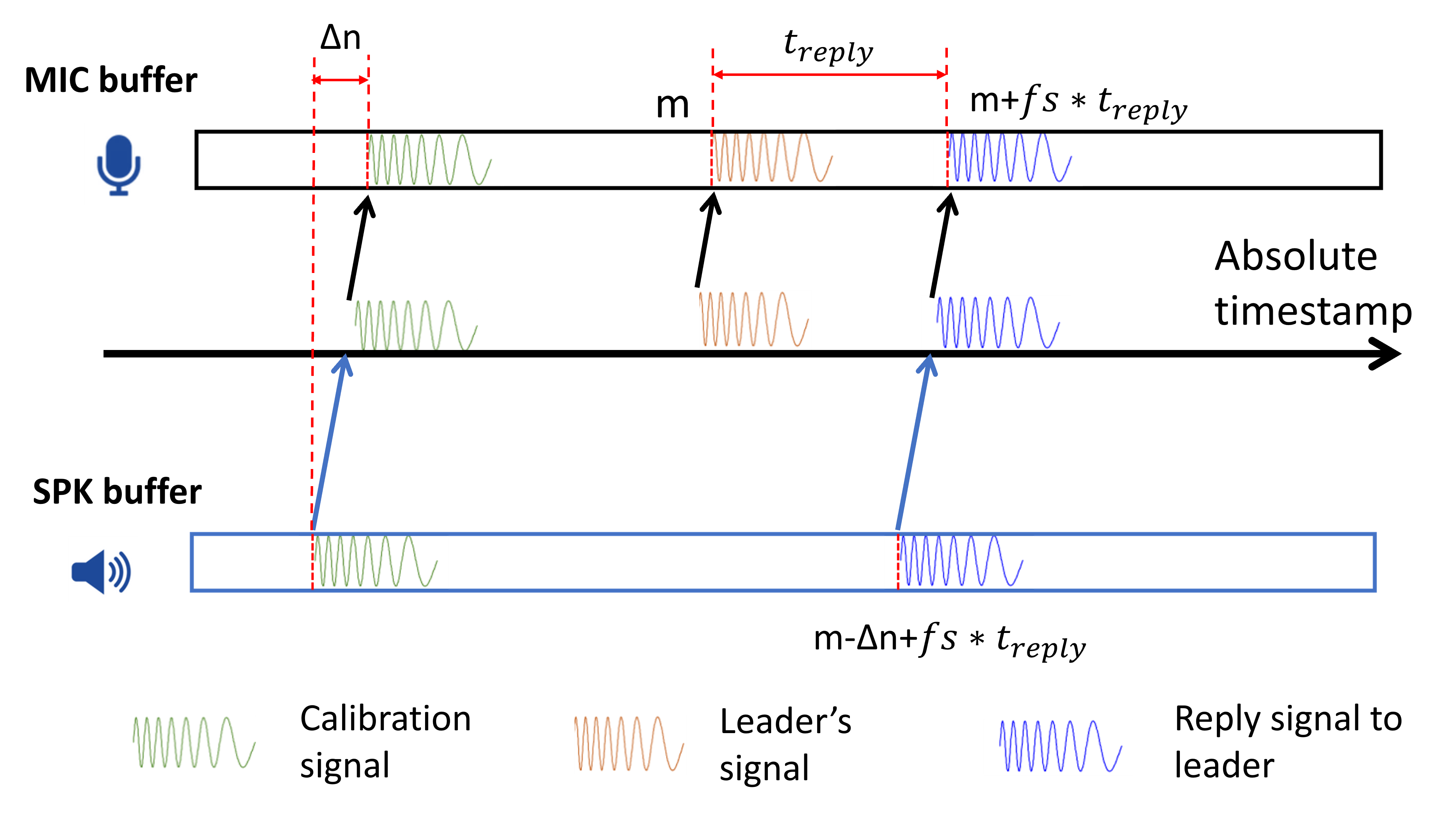}
    \vskip -0.15in
\caption{{\bf Synchronization of microphone \& speaker buffers.}}
\vskip -0.15in
\label{fig:pipeline1}
\end{figure}

In our design, the dive leader initiates the protocol by transmitting a short message with a preamble and its ID and the other devices respond to it using time-division multiplexing (TDM).  Each device is  pre-assigned a unique ID, where the leader device is assigned the ID 0, and other user IDs are from 1 to N-1.
After all devices respond, the timestamp from each user can be used to compute the pairwise propagation time and  distances. We define the propagation time between device $i$ and $j$ as $\tau_{ij}$. The distance between them can be computed as $D_{i,j} = c\tau_{ij}$.
 Since  no global synchronization clock exists underwater,  device $i$ records its local time $T^i$. We define $T^i_j$ as the time when the message from device $j$ arrived at the microphone buffer of device $i$. 


\vskip 0.05in\noindent{\bf  All devices are in leader's  range.} 
All  devices use the leader's message to synchronize and respond in a  TDM fashion. When device $i$ receives the leader's message, it sets its local time to 0 i.e., $T^i_0 = 0$. 
Then each device  divides its  local time into slots and responds in the slot  ordered by their ID. Specifically, device $i$ will send its  message containing a preamble and its ID at $T^i_i = \Delta_0 + (i-1)\Delta_1$ based on its  local clock. 
$\Delta_0$ is the maximum time it takes to process the message from the leader in real-time  as well as the smartphone audio input and output latency. $\Delta_1 = T_{packet} + T_{guard} $, where $T_{packet}$ is the  duration of the message and $T_{guard}$ is the guard interval which accounts for the maximum propagation delay to avoid packet collisions. When all devices are in leader's range,  $t_{guard}$ should be larger than twice  the maximum possible propagation time within the diver group $\tau_{max}$ (i.e. $T_{guard} > 2\tau_{max}$) to guarantee no packet collisions.

\begin{figure}[t!]
    \includegraphics[width=.45\textwidth]{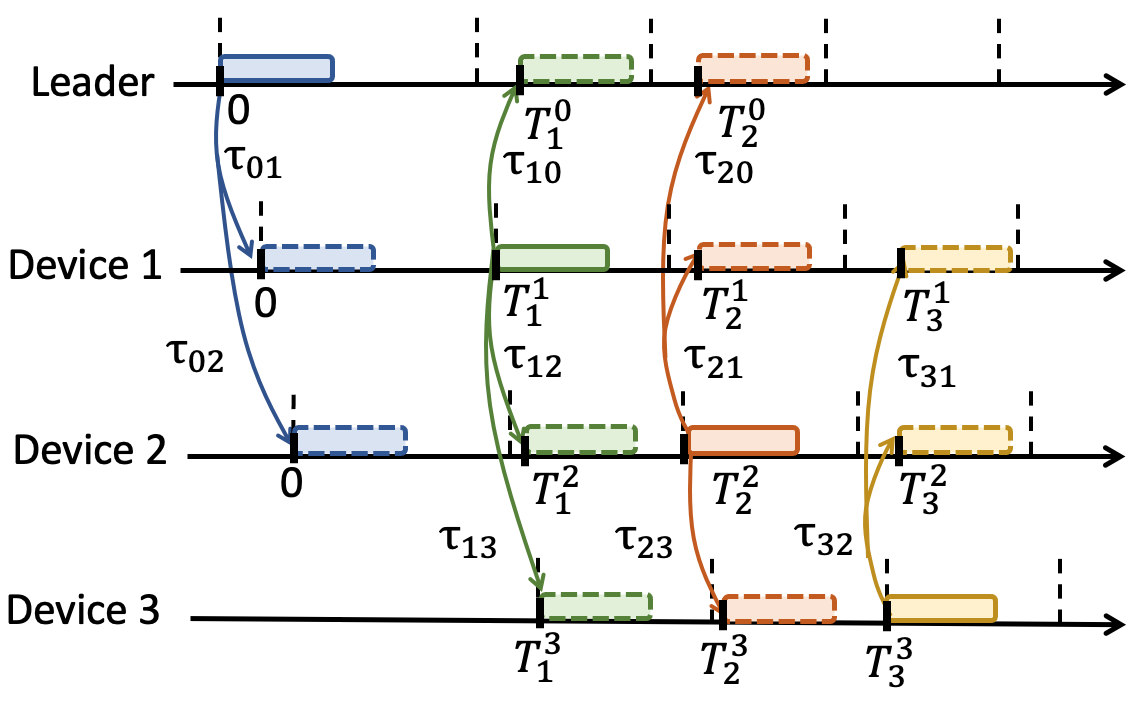}
    \vskip -0.15in
\caption{{\bf Distributed timestamp protocol.}}
\vskip -0.15in
\label{fig:protocol}
\end{figure}


The protocol stops after $N$ slots which is the number of devices in the network. At the end of this protocol, each device records the timestamps at which they received messages from all devices in their range and transmit this timestamp information to the leader, as described in~\xref{sec:comm}. Given these timestamps, the leader can compute the pairwise distances between all devices. Specifically, the distance between device $i$ and $j$ ($i < j$) can be computed as follow:
$D_{ij} = \frac{c}{2}[(T^i_j - T^i_i) - (T^j_j - T^j_i)].$
{Here, we ignore the propagation time from the device's speaker to its own microphone, because it is small compared to underwater distances.}

\begin{figure*}[t!]
    \includegraphics[width=0.95\textwidth]{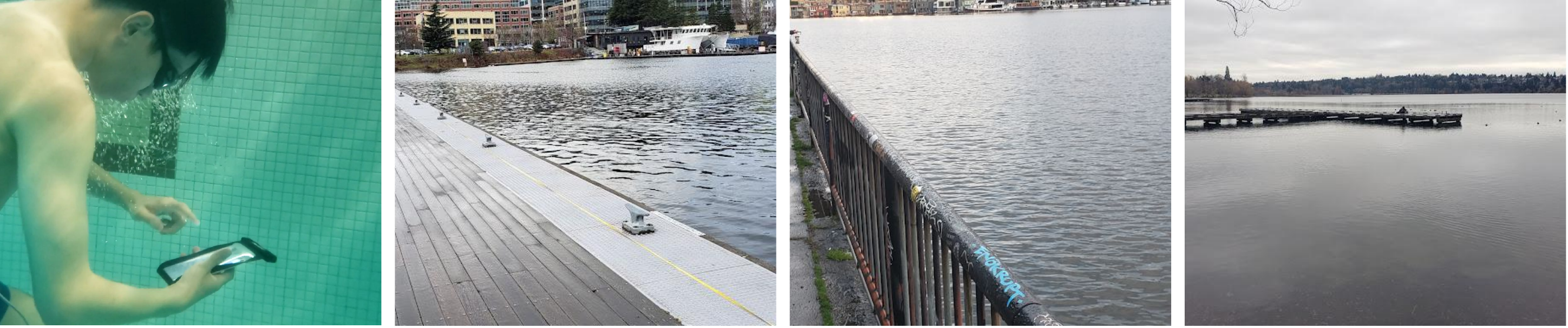}
    \vskip -0.15in
\caption{Different underwater  scenarios. {(a) Swimming pool, (b) Dock, (c) Viewpoint, (d) Boathouse.}}
\vskip -0.1in
\label{fig:locs}
\end{figure*}

\vskip 0.05in\noindent{\bf Not all devices are in leader's  range.} 
As before, the subset of devices that receive the leader's message respond in their assigned TDM  slots. 
Say device $i$ did not receive the leader's message but received a message from other devices. In this case, device $i$ uses the first message it received to synchronize its local time and compute its assigned transmission time slot (Fig.~\ref{fig:protocol}). 
Here, we have two cases. 
 {First, say device $i$ received a message from device $j$ (where $ (i - j)\Delta_1 > \Delta_0$) at time $T^i_j$. It estimates its transmission timeslot as  $T^i_i = T^i_j + (i-j)\Delta_1$. In its  slot, device $i$ transmits its ID and the ID for device $j$ to ensure that  other devices know that its transmit time was set with respect to  device $j$. 
Next, say the first message received by device $i$ is from a device with an ID $j$ (where $(i - j)\Delta_1 < \Delta_0$). In this case, it will miss its slot  and  will have to wait for all remaining devices to transmit before it has a chance to respond in the $T^i_i = T^i_j + (N - j + i)\Delta_1$ time slot. }  
We note  three points.
\squishlist
\item {\it Packet losses.} If there is some device $k$ such that both device $i$ and $j$ received its message, then we can compute the  distance between $i$ and $j$ even if one of the messages from either $i$ to $j$ or $j$ to $i$ is lost.
\item  {\it ID encoding.} We  use MFSK  to encode the ID. We divide the 1-5~kHz frequency band into $N$ bins  ($N$ is the dive group size). For user $i$, we  set all bins of $i^{th}$ bin to 1 and all other bins to 0. We use a maximum-likelihood estimator to  decode the user ID. 

\item{\it Latency analysis.} We set  $\Delta_0 = 600~ms$, $T_{packet} = 278~ms$, $T_{guard} = 42~ms$, and $\Delta_1 = 320 ms$. When all divers are in the leader's range, the maximum round trip time for a protocol run  is $T_{round} = \Delta_0 + (N-1)\Delta_1$. When some divers are out of range of leader, in the worst case, the maximum round trip time  is $T_{round} = \Delta_0 + 2(N-1)\Delta_1$.

\squishend

\subsection{Communication system}\label{sec:comm} 
After the distributed protocol, the users need to send the timestamp and depth information to the leader. Due to the limited bandwidth, we need to compress this data. We discretize  depth at a 0.2~m resolution and so we need 8 bits to represent depths between 0-40~m. For timestamps, instead of transmitting the absolute timestamp $T^i_j$, we transmit the time difference between  $T^i_j$ and the assigned time slot for device $j$, i.e. $\Delta_0 + (j-1)\Delta_1$. This time difference is bounded by $[0, 2\tau_{max})$.  We set  $2\tau_{max} = 42ms$  which corresponds to a  maximum propagation distance  of 32~m. For fs=44100Hz, this  is around 1852 samples in  the microphone buffer. At a  2 sample resolution,   the timestamp differences require $log_2(1852/2) \approx 10 bits$. Hence,  with $N$ divers, each device sends $10(N-1) + 8$ bits  to the leader.

We use  FSK which is a widely deployed modulation~\cite{sigcomm22,van2014performance, sui2009evaluation, li2009bit}. The devices transmit simultaneously to the leader device to reduce latency. To do this, we  divide the 1-5~kHz bandwidth into $N$ bands and pre-assign each device to a different band. Device $i$ uses FSK within its band. We apply 2/3 convolutional coding to the payload. We note that this communication system takes around 0.9, 1 and 1.2~s when $N$ is 6, 7 and 8 at a bit rate of 100~bps per device.

When some users are out of range of the leader, they cannot directly send the message back. Thus, a multi-hop communication mechanism is required which is not in the scope of this paper. 



\section{Results}
We evaluated our  system in the four  environments in Fig~\ref{fig:locs}. 

\squishlist
\item \textit{Swimming pool.} The length of the water here is around 23~m. The depth of the swimming pool varied from  1 to 2.5~m. 
\item \textit{Dock.} This outdoor location has a length of around 50~m with a depth of 9~m. Boats and seaplanes would frequently sail or dock at this location with aquatic plants and animals.
\item \textit{Viewpoint.} By the waterfront of a park with a length of 40~m. The water had a depth of around 1 to 1.5~m. 
\item \textit{Boathouse.} Fishing dock by the lake with a horizontal distance of 30~m. The lake had a depth of 5~m. This is a busy location with people fishing and kayaking close to the dock.
\squishend

\subsection{Benchmark evaluation}

\vskip 0.05in\noindent{\it Accuracy versus device separation.} 
Here, we  evaluate  our system along the dock of a lake with an average water depth of 9~m. We performed  experiments using two Samsung Galaxy S9 phones set to transmit at the maximum speaker volume. The phones were set to transmit using the speaker at the bottom of the device, and receive using the microphones at the bottom and top of the device. The experiment was repeated in each location every six seconds. At each distance, the sender and replier are set to exchange messages up to a maximum of 60 times. The measurements were divided into roughly three sessions, where after 20  measurements, the phones were removed  and submerged again. 
The phones were enclosed in a  pouch (Hiearcool waterproof phone pouch~\cite{pouch}) and attached to a selfie stick and telescopic extension pole, which was used to submerge the phones to a depth of 2.5~m. The selfie stick and extension pole were attached using waterproof tape and zip ties. This setup was chosen so that the phone's position and angle could be controlled. We used a tape measure to mark   distances up to  45~m. 


Fig.~\ref{fig:main_result2}a shows the CDF of the absolute error obtained by our system for four distances of 10, 20, 35 and 45~m.  The error in distance increases with separation between the devices because the signal strength is lower at larger separation.  We also analyze the effect of using both the top and bottom microphones for ranging versus using only a single microphone in isolation. Fig.~\ref{fig:main_result2}b shows the 95th percentile distance error for these scenarios at distances of up to 45~m. The figure reveals the following: firstly, utilizing both microphones yields lower ranging errors at all distances. This can reduce error by as much as 4.52~m at a distance of 45~m \textcolor{red}{(while we set the maximum distance to 32m in the distributed protocol, here we relax this to evaluate the limits of 1D localization)}. Secondly, when a single microphone is used in isolation, there is no clear relationship between microphone position and  error. 





\begin{figure}
    \includegraphics[width=.23\textwidth]{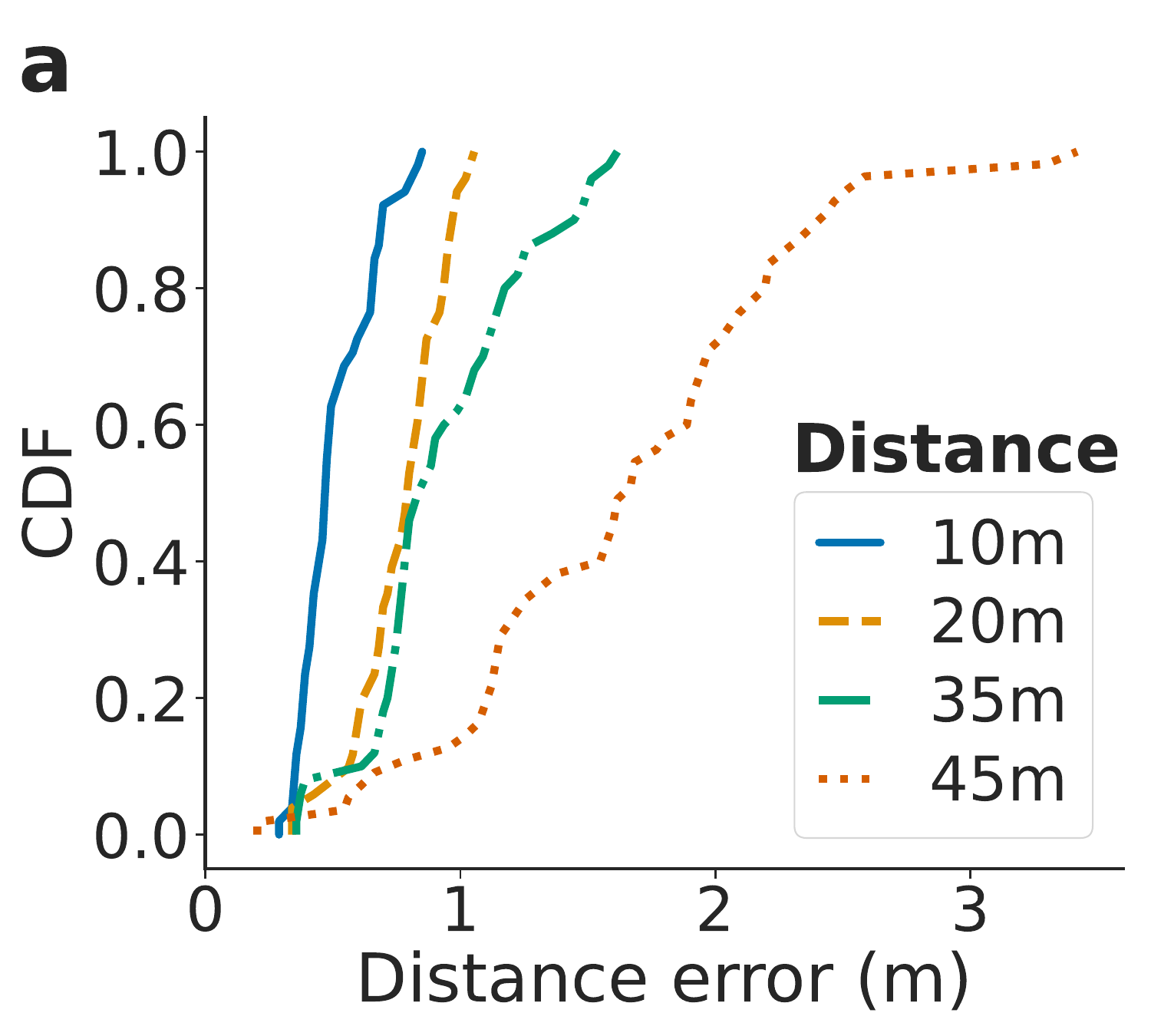}
    \includegraphics[width=.23\textwidth]{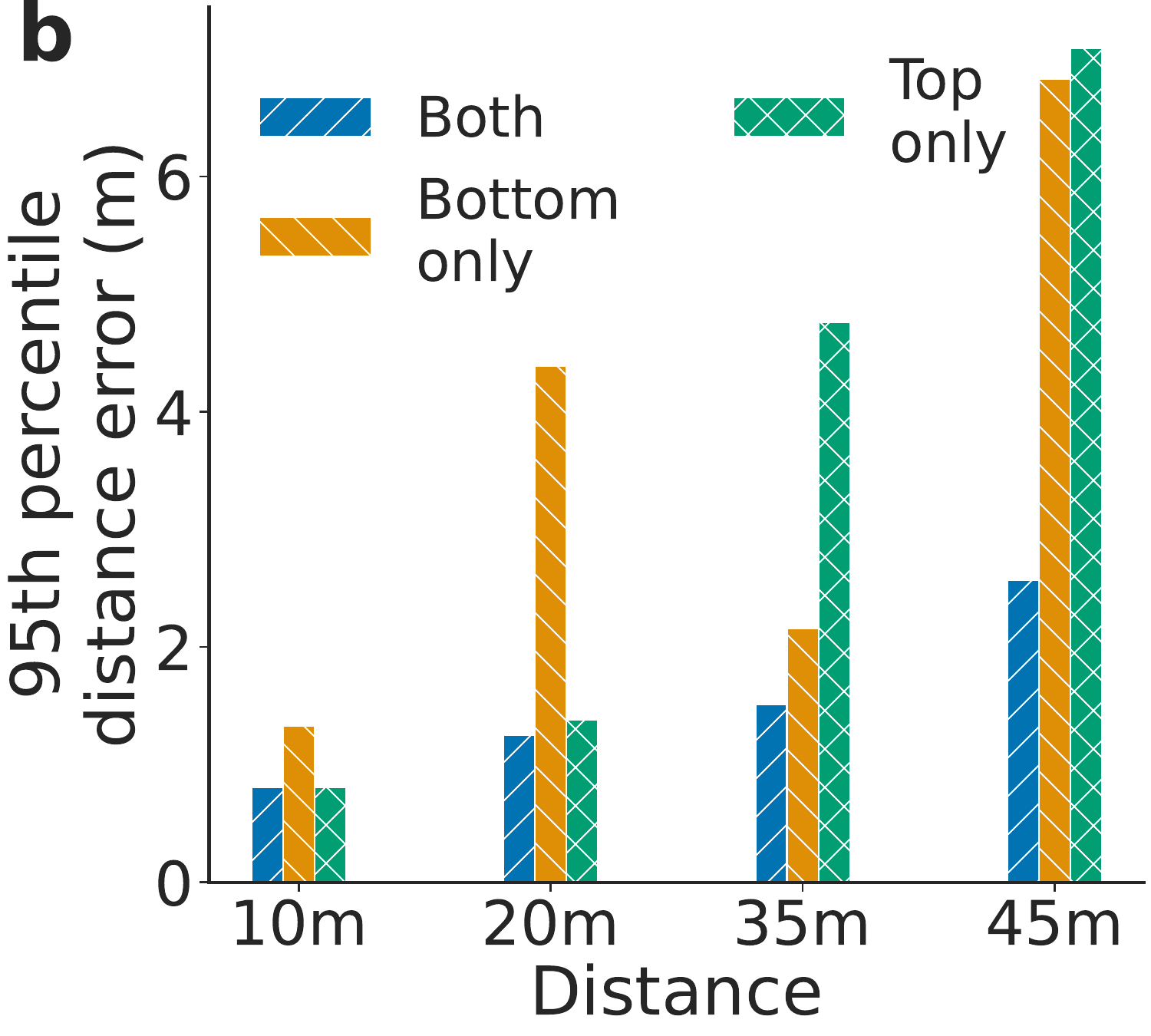}
    \vskip -0.15in
\caption{(a) Ranging accuracy v/s  separation. CDF of absolute error as a function of separation.  (b) 95\% errors using both microphones, the bottom and top microphone only.}
\vskip -0.15in
\label{fig:main_result2}
\end{figure}

\begin{figure}[t!]
    \includegraphics[width=.23\textwidth]{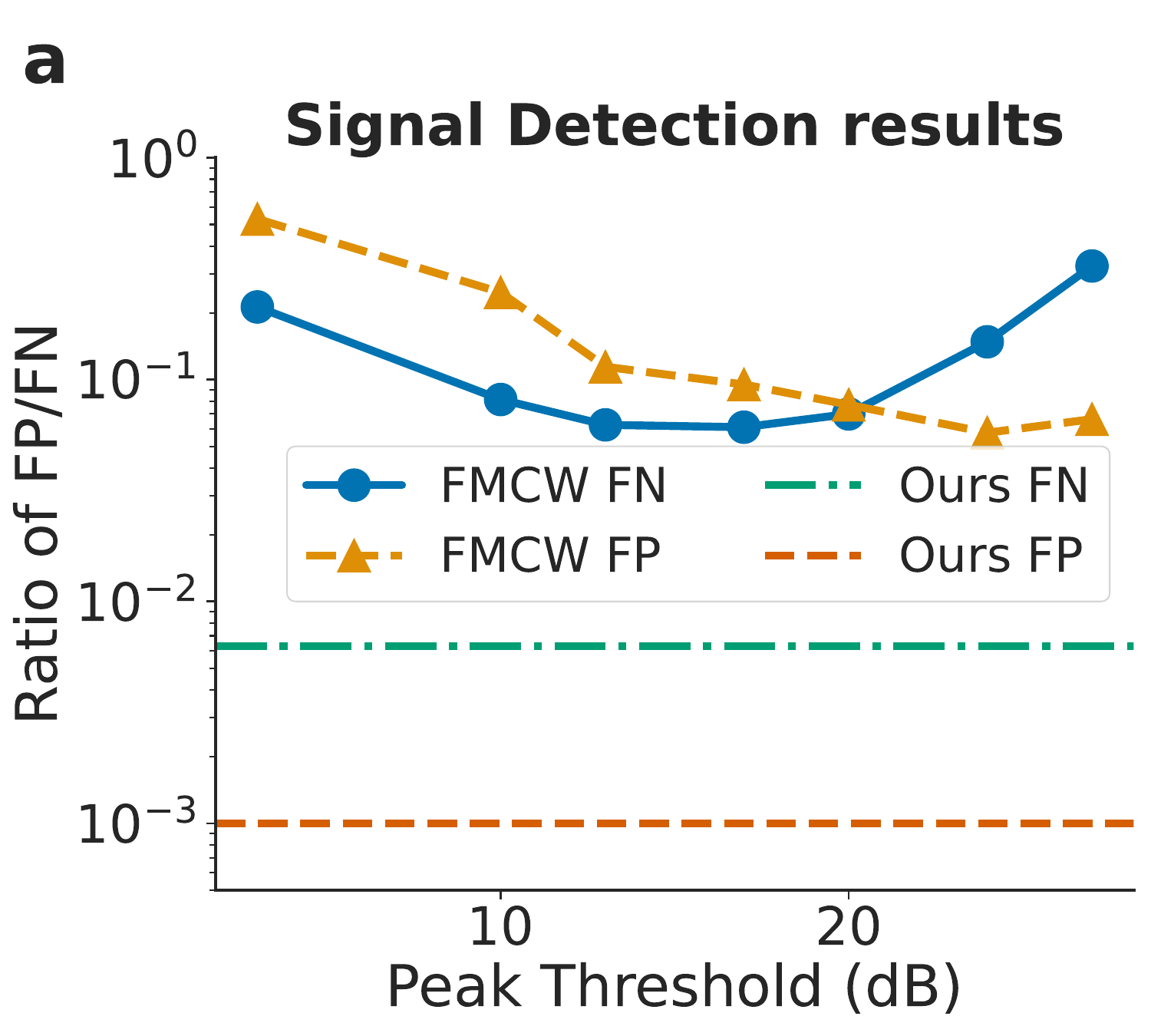}
    \includegraphics[width=.23\textwidth]{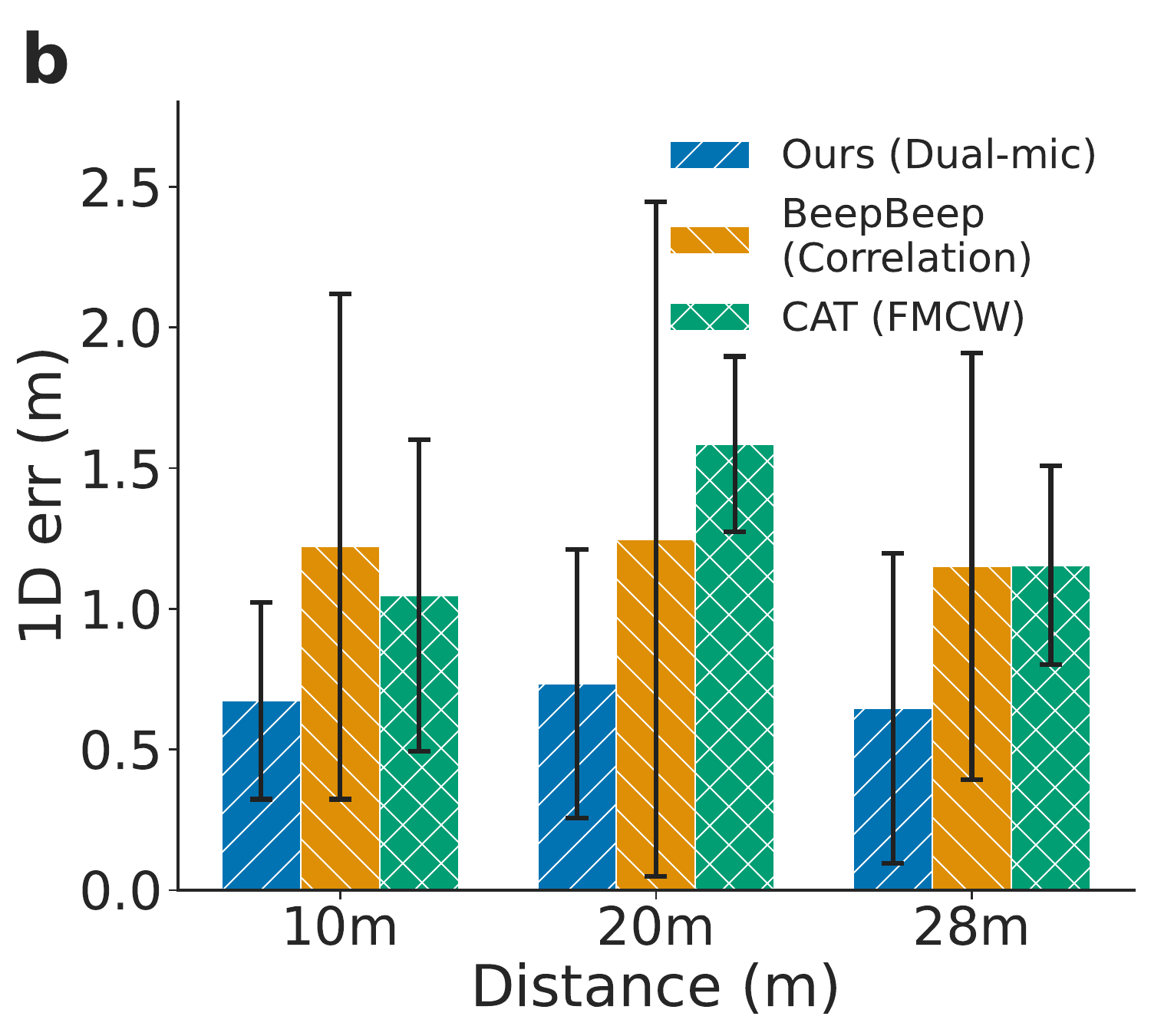}
    \vskip -0.15in
\caption{\textcolor{red}{(a) Ratio of false positive and false negative for signal detection  (b) 1D ranging error versus distance.}}
\vskip -0.15in
\label{fig:main_result3}
\end{figure}

\tuochao{We also compare our 1D ranging algorithm with previous works on acoustic-based 1D ranging and tracking like~\cite{peng2007beepbeep, mao2016cat}.  \cite{peng2007beepbeep} uses the linear chirp signal and applies auto-correlation with specially-designed peak detection. \cite{mao2016cat} implements an FMCW-based approach where the receiver mixes the received signal with the transmitted signal. For a fair comparison, we control the duration and bandwidth of three types of signals to be the same. 
We put the phones at horizontal distances of 10, 20, 28m with a depth $\sim 1m$ at the boathouse location. 
To compare the robustness of signal detection, we send three types of preambles for 180 times at each distance. We implement the window-based power threshold $TH_{SD}$ in ~\cite{peng2007beepbeep} to detect the FMCW signal. Since $TH_{SD} = 3dB$ in ~\cite{peng2007beepbeep} is selected for in-air experiment, we try different $TH_{SD}$ to calculate the false positive and false negative for fair comparison. Fig.~\ref{fig:main_result3} (a) shows that our preamble detection is more robust than the state of art.
To compare the 1D ranging algorithm with ~\cite{peng2007beepbeep, mao2016cat}, the phones exchange each of different signals $\sim 60$ times. Then we apply our dual-mic channel estimation, auto-correlation~\cite{peng2007beepbeep}  and  FMCW\cite{mao2016cat} on these measurements to calculate the 1D ranging error.
Fig.~\ref{fig:main_result3} (b) shows the mean value and standard deviation of the 1D ranging errors demonstrating that our approach outperforms prior work}.

\vskip 0.05in\noindent{\it Accuracy versus depth.} 
We place the smartphones  at different water depths at the dock location which had a total depth of 9~m.  We lowered the smartphones into the water using ropes marked at  2, 5, and 8~m. The phones were weighed down with a bag of pebbles to ensure the ropes were vertical. The phones were positioned at a  horizontal distance of 18~m. Unlike previous experiments, the rope would cause the phone to rotate and sway slowly.  We repeat  measurements thrice at each depth. Fig.~\ref{fig:depth_dual}a shows that the median and 95th percentile error is lowest at 0.28 and 0.73~m for the 5~m depth, which is around the midpoint depth of the dock.  This likely is because multipath reflections  can be stronger when the devices are close to the surface or floor of the water body. 


\begin{figure}[t!]
\includegraphics[width=.23\textwidth]{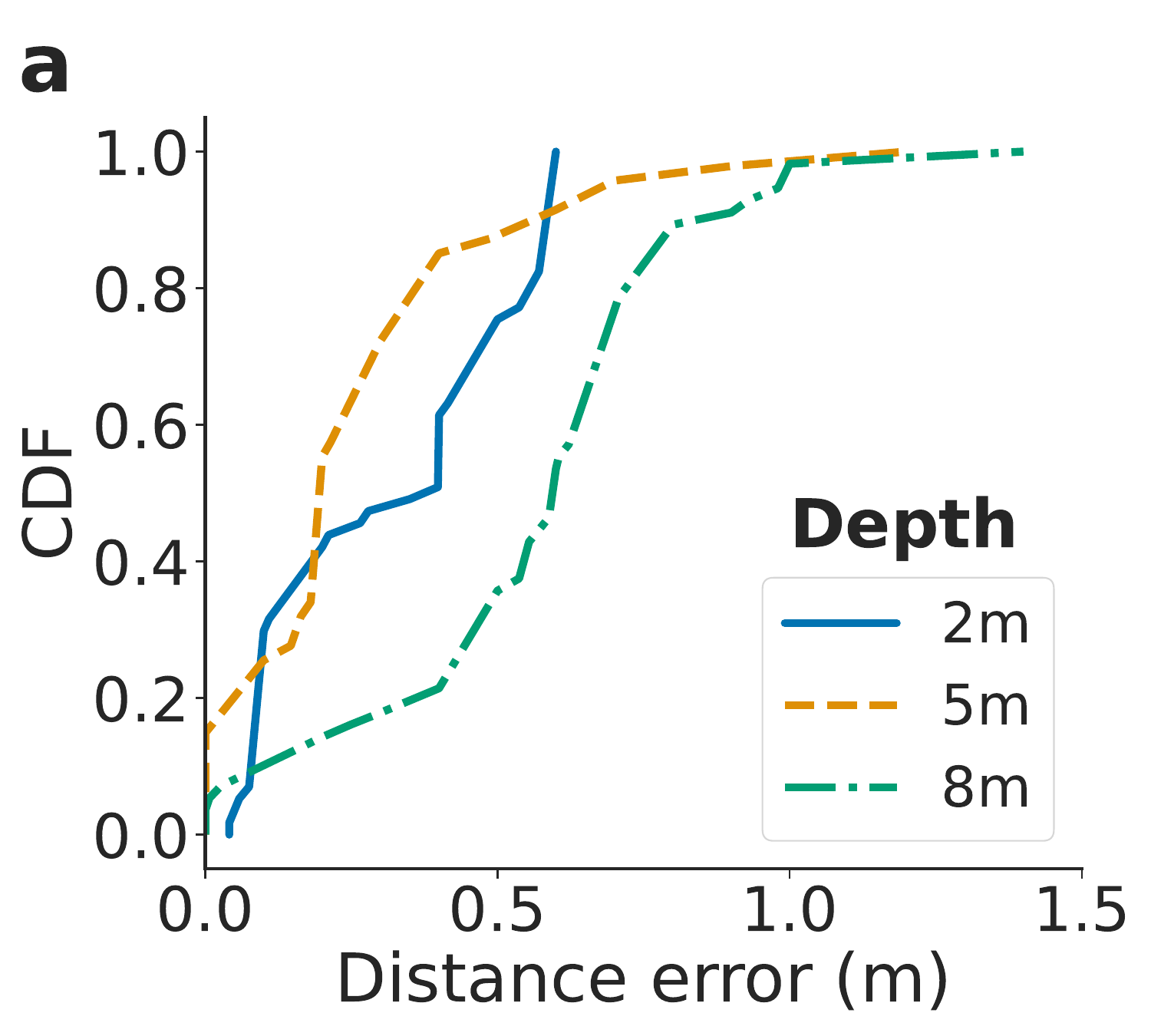}
    \includegraphics[width=.23\textwidth]{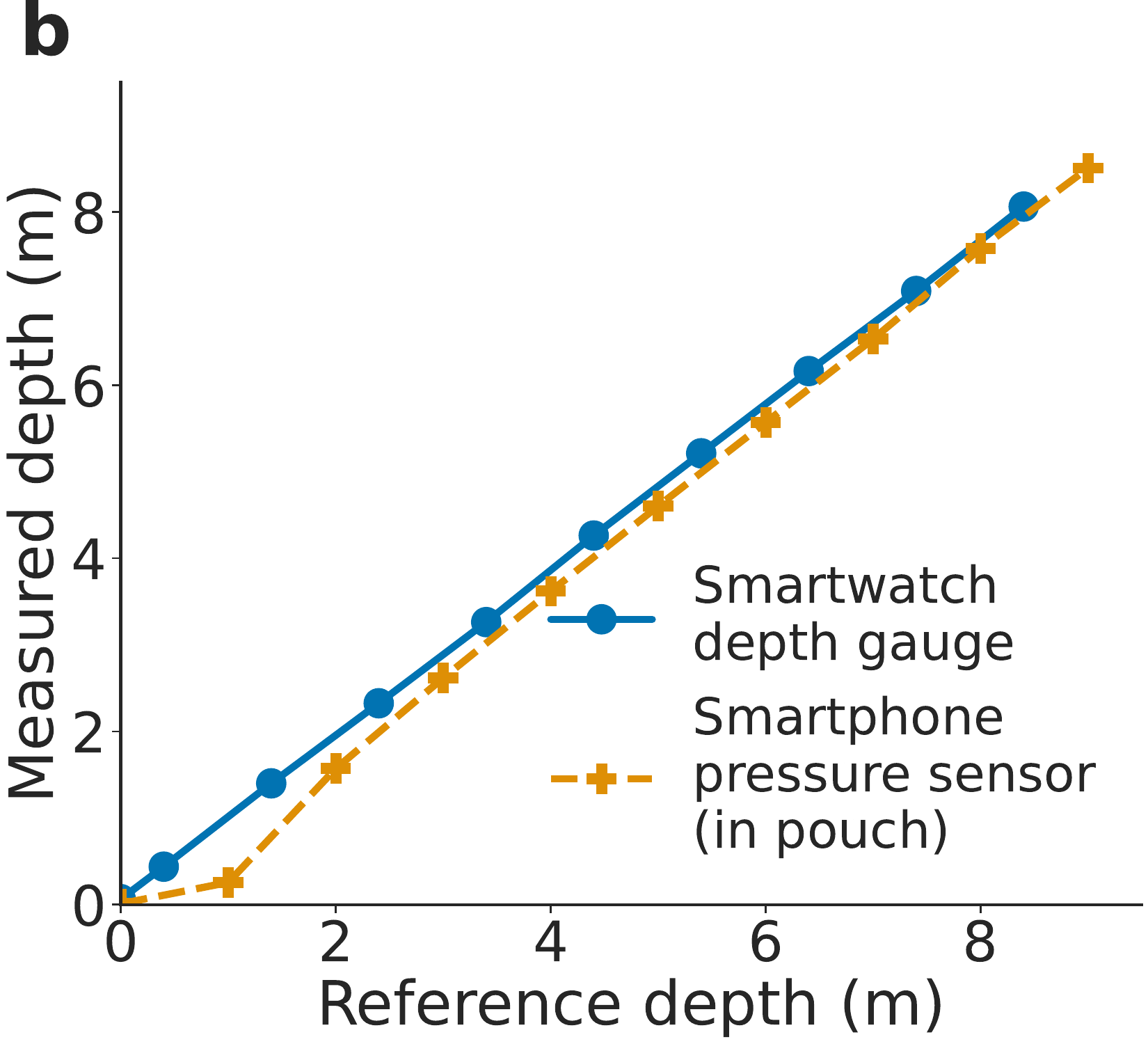}
\vskip -0.15in
  \caption{(a) Effect of device depth. Errors for different depths with  devices  separated horizontally by 18~m. (b) Accuracy of depth measurements from smartwatch and phone. }
  \vskip -0.1in
  \label{fig:depth_dual}
\end{figure}

\begin{figure}[t!]
    \includegraphics[width=.23\textwidth]{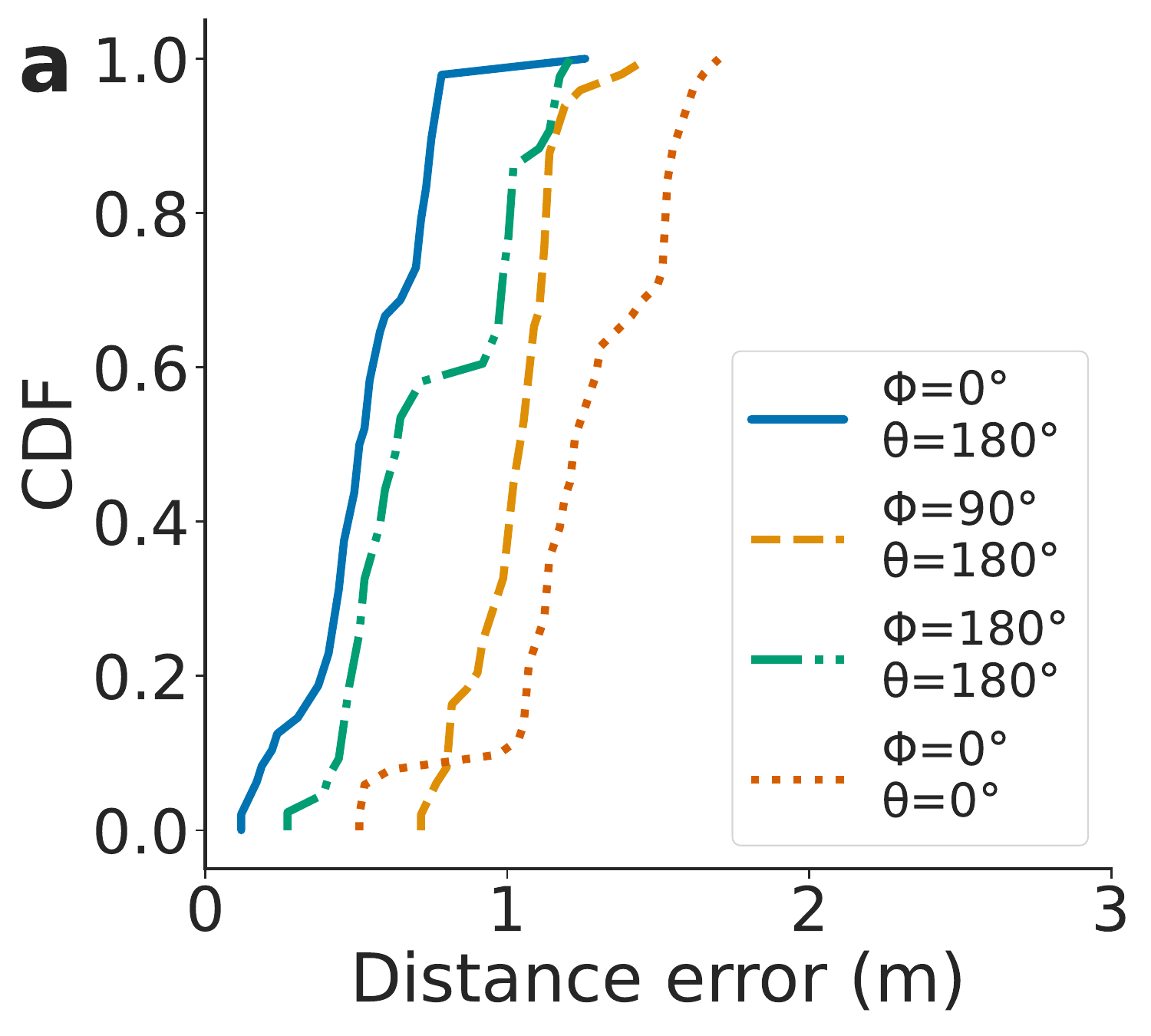}
    \includegraphics[width=.23\textwidth]{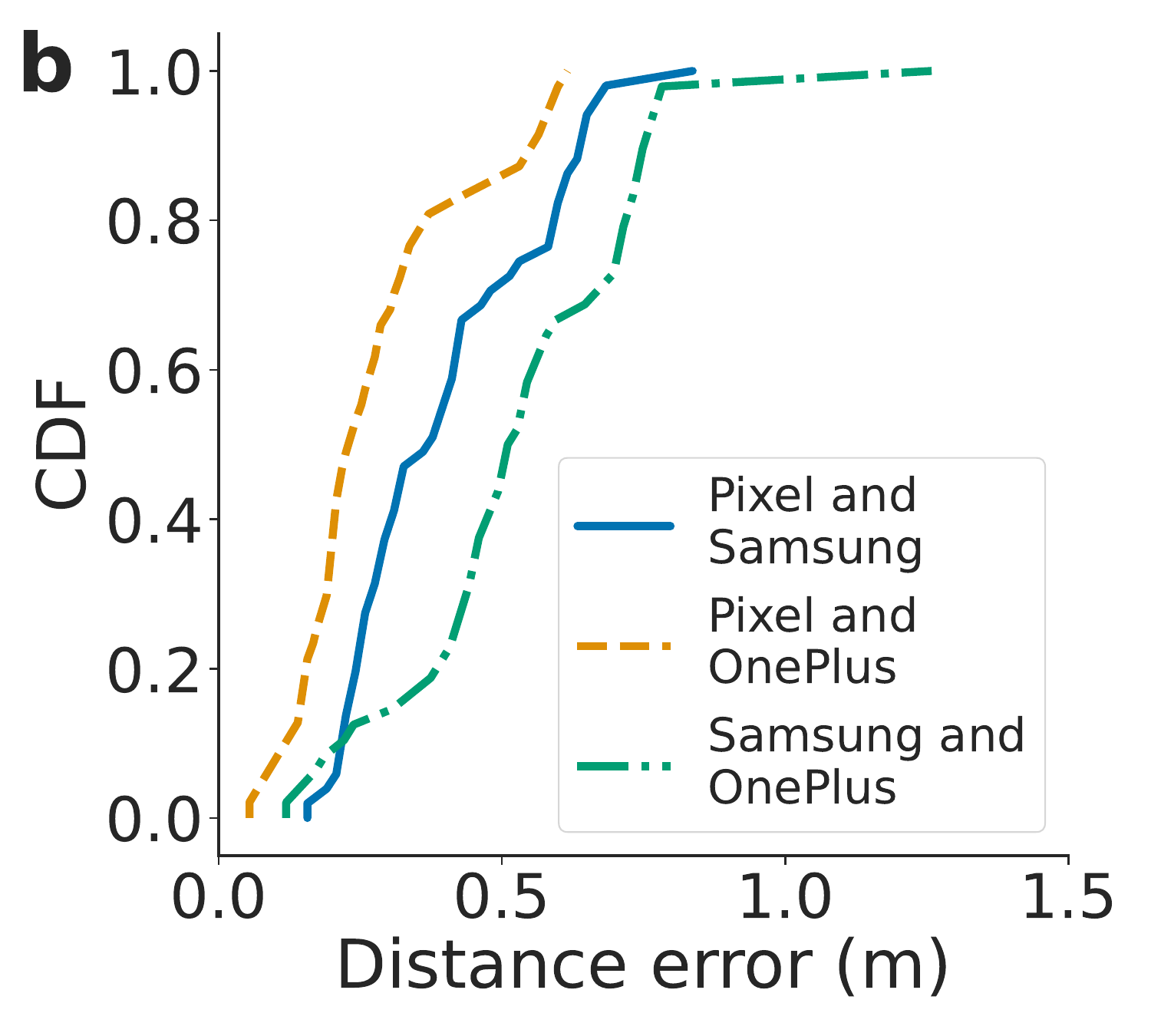}
    \vskip -0.15in
\caption{Effect of (a) orientations with phones separated by  20~m, and (b) various   smartphone model pairs.}
\vskip -0.1in
\label{fig:phones_rot}
\end{figure}

\vskip 0.05in\noindent{\it Effect of phone orientation and make.} 
We evaluate this at the dock at a horizontal distance of 20~m and a depth of 2.5~m. We first positioned the speaker and microphone of both phones to directly face each other so their azimuth $\phi$ and polar angle $\theta$ is set to 0° and 180° respectively. We measure the  error when the sender phone is rotated to different azimuth and polar angles. We first  rotate the sender phone in the azimuth angle to 90° and 180° while keeping the polar angle constant. We then reposition the phone so its speaker and microphone faces upwards with  $\phi=0^{\circ}$, $\theta=0^{\circ}$. Fig.~\ref{fig:phones_rot}a shows that median  error ranges from 0.54 to 1.25~m. Note that when the phone faces upwards it had the largest error likely because the phones are closer to the water surface  resulting is higher multipath when pointing towards the surface. We also evaluated our system with different smartphone model pairs. In Fig.~\ref{fig:phones_rot}b, we evaluated three different Android  phones models. 

\vskip 0.05in\noindent{\it Depth accuracy.}\label{sec:depth} We evaluate the  depth gauge sensor on the Apple Watch Ultra and use the pressure sensor on the Samsung Galaxy S9 smartphone in a waterproof case for estimating underwater depth. The depth gauge readings from the smartwatch were obtained from the Oceanic+ app. The smartphone's pressure sensor values $P$ in units of Pascals are converted to depth measurements $h$ in units of meters using the equation~\cite{pressure_equation}: $h=\frac{P-P_0}{\rho g}$, where $\rho=997~kg/m^3$ is the average density of water, $g=9.81~m/s^2$  and $P_0=101325~Pa$ is atmospheric pressure  at sea level. We performed this evaluation in the dock location which had a depth of 9~m, and lowered each smart device underwater with a rope in increments of 1~m using markings on the rope as a measure of ground truth. The devices were  held in place at each depth for 30 s. Fig.~\ref{fig:depth_dual}b shows the accuracy for depth from the smartwatch and smartphone. Across all measurements, the average depth error on the smartwatch and smartphone were $0.15 \pm 0.11~m$ and $0.42 \pm 0.18~m$, respectively.

\begin{figure}[t!]
    \includegraphics[width=.23\textwidth]{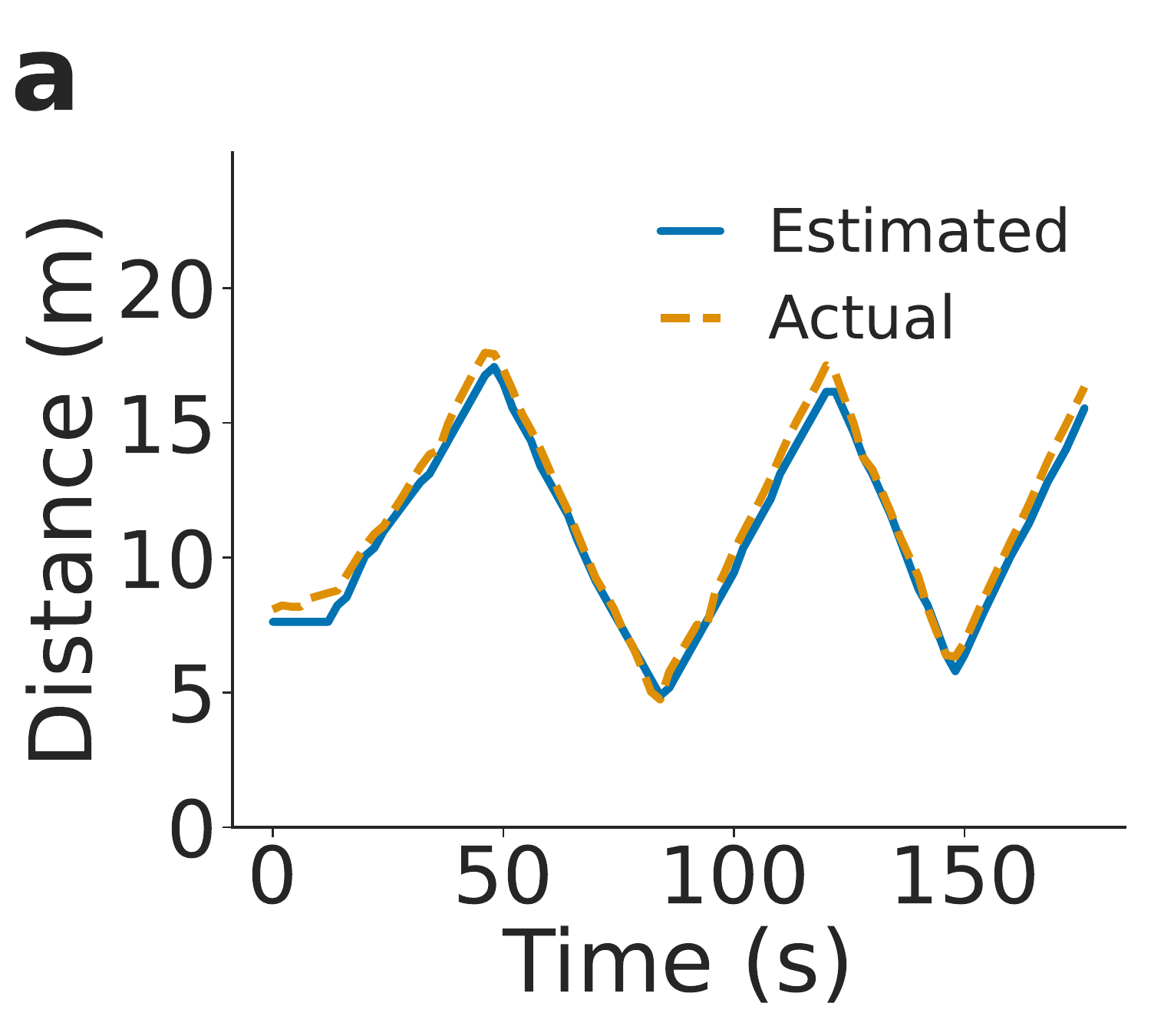}
    \includegraphics[width=.23\textwidth]{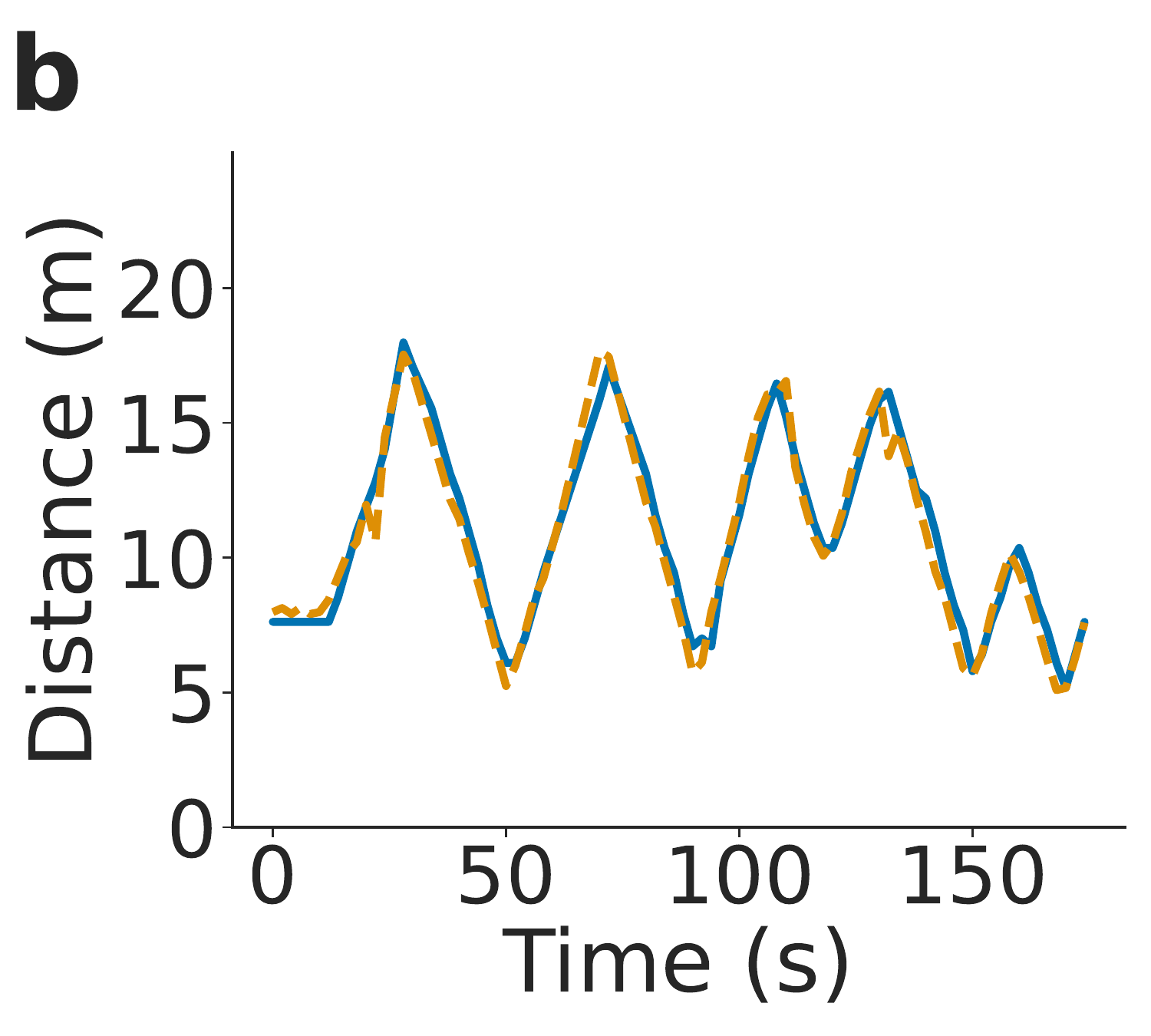}
    \vskip -0.15in
\caption{\textcolor{red}{1D Ranging errors for  moving devices.}}
\vskip -0.15in
\label{fig:motion1d}
\end{figure}

\begin{figure}[t!]
    \includegraphics[width=.48\textwidth]{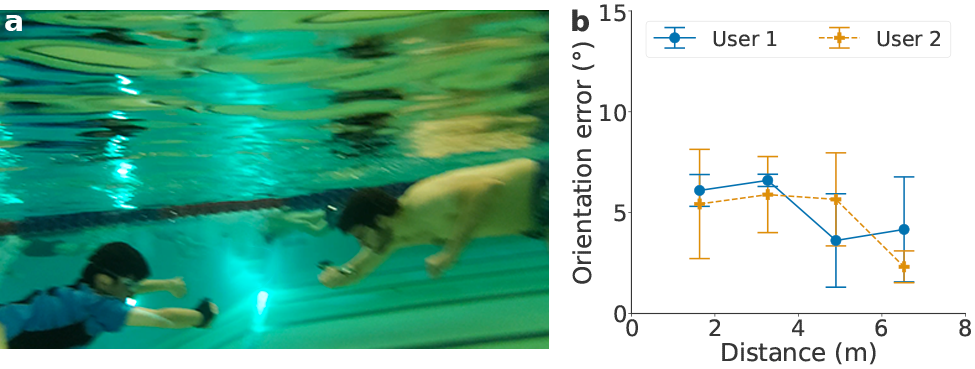}
    \vskip -0.15in
\caption{Human orientation evaluation.}
\vskip -0.15in
\label{fig:orientation}
\end{figure}

\vskip 0.05in\noindent{\it Effect of motion.} 
\tuochao{We first evaluate how our 1D ranging algorithm works when the device keeps continuously moving. In this experiment, we keep a phone static and attached another phone to the extension pole in the dock location. The transmitter sends the preamble every 1 second. Then we moved the extension pole along different 1D trajectories parallel to the coast. To obtain the ground-truth trajectory, we placed a measurement tape along the dock coast, mounted a camera on the extension pole, and pointed it to the measurement tape. Since the movement of the phone on the extension pole is parallel to the measurement tape, the ground truth of distance changes between the two phones can be recorded by the camera video (yellow lines in Fig.~\ref{fig:motion1d}). 
As shown in Fig.~\ref{fig:motion1d}a,b, the smartphone was moved at an average speed of 32 and 56 cm/s.  The median and 95th percentile 1D  errors were 0.51 and 1.17~m.}

\vskip 0.05in\noindent{\it Leader orientation accuracy.}  We evaluate the ability for a dive leader to orient themselves to face a diver within their visual range. To do this, we measure the orientation error with respect to two users in a swimming pool at different distances. Both users wore a wristband (HCcolo Wristband Phone Holder) that held a smartphone in a waterproof case. One user (the diver) stayed in a stationary position at the end of the pool with their smartphone and also held a $4 \times 5$ checkerboard pattern used for camera calibration. The other user (the leader) was positioned at a fixed distance initially at a random orientation, and would then rotate their body and arm to directly face the stationary user. The leader's smartphone was set to capture video footage of the orienting motion, and the process was repeated at different distances.  To compute the orientation error, we leverage algorithms from prior work~\cite{camera_matlab} to calculate the  position in world coordinates of the camera $C=(x_{C},y_{C},z_{C})$, the stationary user's checkerboard/phone $P=(x_{P},y_{P},z_{P})$, and the center of the camera frame $D=(x_{D},y_{D},z_{D})$. We define the vector between the camera and the checkerboard/smartphone as $\vec{v}_{PC}=P-C$ and the vector between the camera and the center of the camera frame as $\vec{v}_{DC}=D-C$. We then compute the orientation error as the angle between the two vectors $cos^{-1}(\frac{\vec{v}_{PC} \cdot \vec{v}_{DC}}{|\vec{v}_{PC}| |\vec{v}_{DC}|})$. If the checkerboard/smartphone is at the center of the image frame, the orientation error would be $0^{\circ}$. Fig.~\ref{fig:orientation} shows our evaluation with two different users as the lead diver. The average error across both users and distances is $5.0^{\circ}$. 



\begin{figure}[t!]
    \includegraphics[width=.48\textwidth]{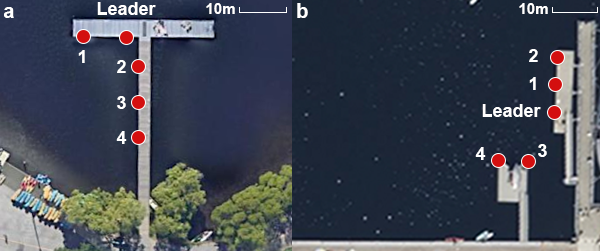}
    \vskip -0.15in
\caption{Testbeds used to evaluate our positioning system at the (a) dock and (b) boathouse locations.}
\vskip -0.15in
\label{fig:topos}
\end{figure}

\begin{figure}[t!]
    \includegraphics[width=.23\textwidth]{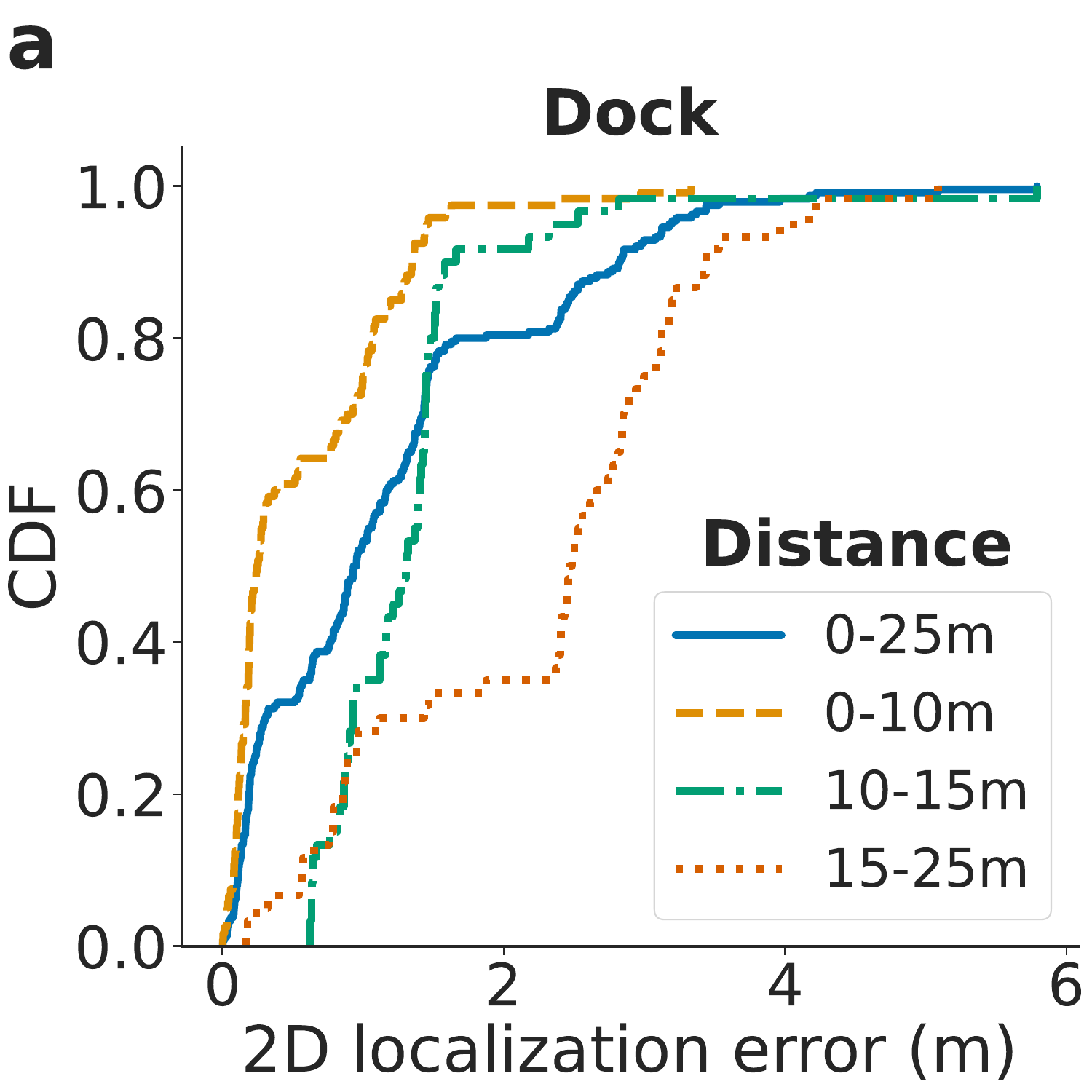}
    \includegraphics[width=.23\textwidth]{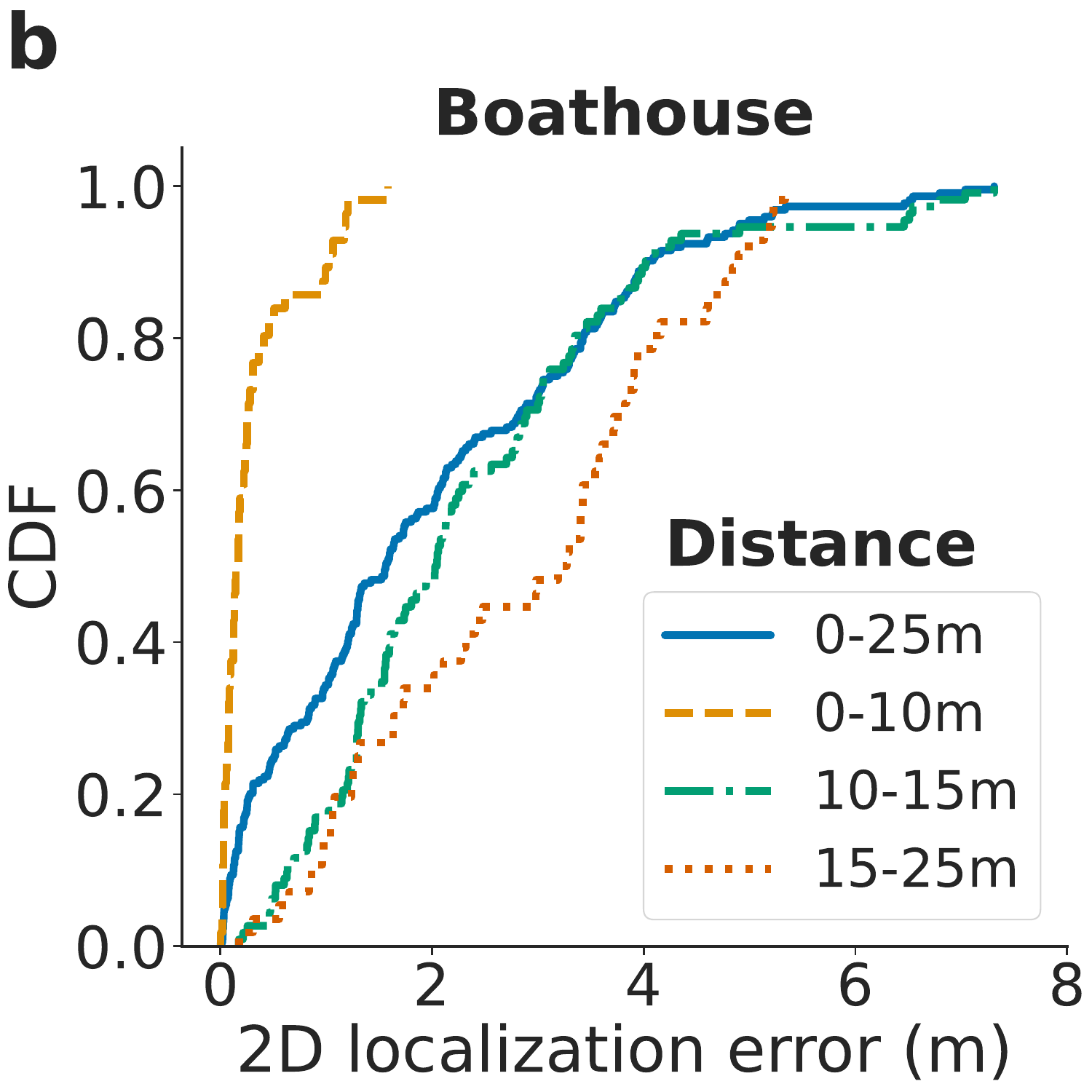}
    \vskip -0.15in
\caption{2D localization error broken down by links of different distances at the (a) dock and (b) boathouse.}
\vskip -0.15in
\label{fig:2dresult}
\end{figure}

\vskip 0.05in\noindent \red{\textit{Battery life.} We evaluate the power consumption of our system on an Apple Watch Ultra and Samsung Galaxy S9 smartphone. To do this, we had the smartwatch continuously play the Emergency SOS~\cite{emergency_sos} siren, and had the smartphone continuously transmit the system preamble at maximum volume every three seconds. The sound levels produced by both devices were 85 and 88~dB SPL respectively at a distance of 1~m in air. The smartwatch and smartphone's battery power reduced by 90\% and 63\% respectively after a duration of 4.5 hours which is longer than the maximum recommended dive times for recreational scuba diving~\cite{padi_table,padi2}.}

\subsection{Network testbed  evaluation}

\vskip 0.05in\noindent{\it 2D localization accuracy versus device separation.} \label{sec:2dloc} We evaluate  our system at the dock and boathouse locations using a network  of five devices as shown in Fig.~\ref{fig:topos}. The topologies were chosen such that the pair-wise distances between smartphone nodes spanned a range of distances from 3 to 25~m from the leader device. Each smartphone was submerged underwater with a rope at different depths. The ground truth for the 2D locations in the dock location was measured using a measuring tape. At the boathouse location, the 2D ground truth was obtained using the distance measuring tool on Google Maps as the  devices were distributed onto two separate islands that were separated by a body of water. In each configuration, a total of {approximately 240 measurements} were collected. Each measurement was split across roughly 5 sessions, where the devices were  retrieved from the water and submerged between measurements. Fig.~\ref{fig:2dresult} shows the CDF of the 2D localization errors for the devices at the dock and boathouse broken down by link distance to the leader device. The median (95\%) errors across all devices were 0.9~m  (3.2~m) and 1.6~m (4.9~m) in the two deployments. As excepted the error increases as the distance to the leader diver increases. This is  acceptable since localizing a device at 20~m to within 3~m,  is sufficient for our target  application.

\vskip 0.05in\noindent{\it Effect of erroneous links.} We evaluate this in the dock location by blocking the link between the leader and user 1 with a thick solid sheet attached to a telescopic extension pole. To ensure the link was completely occluded, we place the leader and user 1 at the same depth of 1.5~m. Note that despite being occluded, the devices can hear each other but have an erroneous distance estimate. The median and 95\% error in this network deployment {with our outlier detection algorithm applied} were 1.4 and 3.4~m respectively. In Fig.~\ref{fig:block_drop}a, we focus on the worse 10\% of the localization errors, i.e., 90--100th percentile. We plot the CDF with and without our outlier detection algorithm applied. The plot shows that without outlier detection, our error estimates have a long tail due to the presence of occlusions. However,  our  outlier detection algorithm  can  address erroneous distance estimates and  reduce  2D localization errors.

\vskip 0.05in\noindent{\it Effect of link removal.}  We use the data collected from the dock location  and  randomly remove a single random link for each of the  measurements.  Fig.~\ref{fig:block_drop}b shows the 2D localization error  with and without link removal. The main observation is that although the median errors between the two scenarios are similar with an error of 1.0 and 0.9~m respectively, the 95th percentile error with link removal is higher than a fully connected network with an error of 6.2 and 3.2~m respectively. The reason for this is that certain links play a greater role in constraining the rotational ambiguities needed to distinguish between nodes that are positioned in a straight line. Dropping these links increases localization error. \tuochao{For  randomly dropping nodes, the localization error does not increase and even become slightly better. This is because some of the far away devices have larger ranging errors. Sometimes dropping the “bad” far-away devices can  improve topology estimation for the remaining devices. } 

\vskip 0.05in\noindent{\it 4-device networks.} We evaluate 4-device network deployments by randomly removing a node   from the network using data from the dock location. Here, we measure the 2D localization error when randomly removing a device  from the network, except for the leader  and user 1. Fig.~\ref{fig:block_drop}b shows that the CDF of 2D localization error of the 5-device and 4-device networks are similar with a median error of 0.9 and 0.8~m, and a 95th percentile error of 3.2 and 3.2~m. 

\vskip 0.05in\noindent{\it Effect of mobility.} \tuochao{We also evaluate whether the motion of the device affects our 2D localization. Specifically, we attach 5 phones to the ropes and place them as shown in Fig.~\ref{fig:topos}. Then we moved a single device forward and backward around its original positions. Since the phone is attached to the rope, its orientation also keeps changing during movement. The speed of motion was between 15-50~cm/s. We performed this evaluation twice, first by moving only user 1 then by moving only user 2. The ground truth for the moving device was set to the midpoint of the trajectory. Our evaluation results in Fig.~\ref{fig:motion2d} show that the change in 2D localization error is modest with the median error for user 1 increasing from 0.2 to 0.3~m when it is moving, while the median error for user 2 increases from 0.4 to 0.8~m when it is moving. This is due to the distributed nature of our protocol which is able to tolerate multi-path variations caused by mobility in the environment.}

\begin{figure}[t!]
    \includegraphics[width=.23\textwidth]{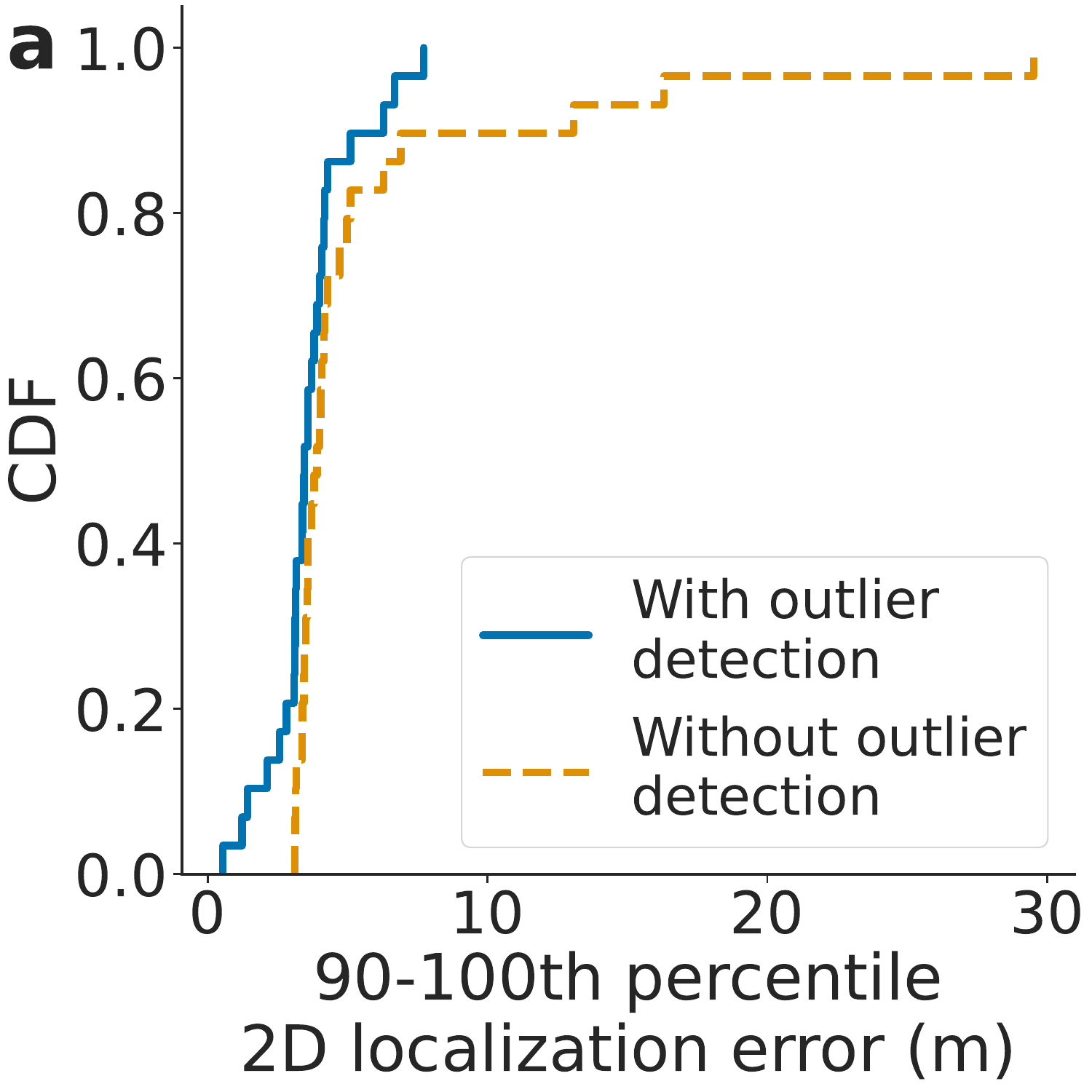}
    \includegraphics[width=.23\textwidth]{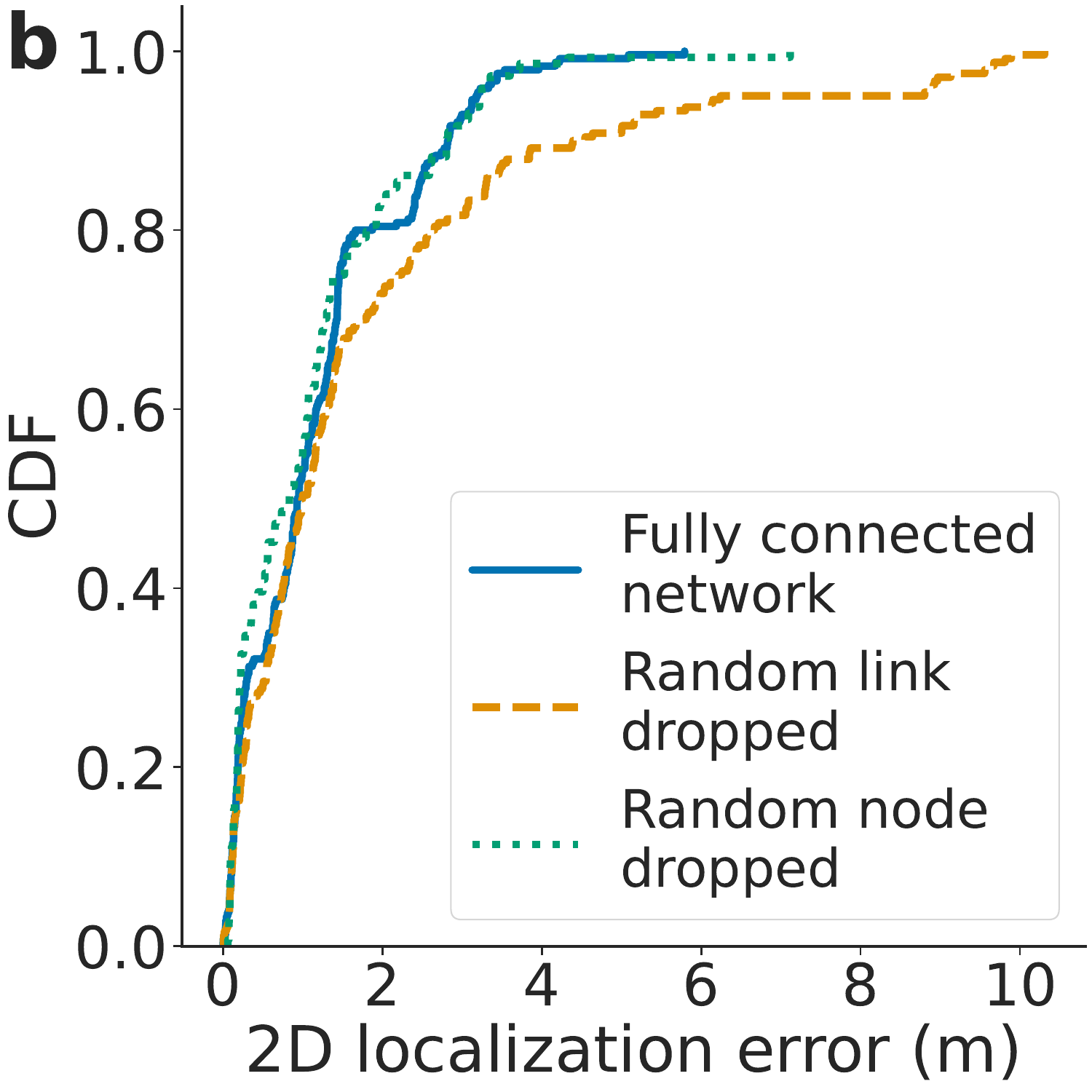}
    \vskip -0.15in
\caption{{Effect of (a) erroneous links due to occlusions, and (b) link and node removal.}}
\vskip -0.15in
\label{fig:block_drop}
\end{figure}

\vskip 0.05in\noindent{\it Flipping disambiguation accuracy.}  We ran  experiments at the dock environment  with 5 devices (Fig.~\ref{fig:topos}(a)). We used a long stick  instead of the rope so that we can control the orientation of the leader device. We  collect a total of 50 sets of localization data points with the leader device oriented towards a closeby device (either device 1 or 2). We run the flipping disambiguation algorithm in 2 settings: (1) we use signals from only one of the 3 devices (excluding leader and the visual device that the leader is pointing to)   to resolve  flipping as described in~\xref{sec:flipping}.  (2) we use the signals from all other 3 devices   to resolve  flipping. Across all 50 experiments, when using only one device's signal the flipping disambiguation accuracy is 90.1\%. When we use all three devices' signals, the flipping disambiguation accuracy improved to 100\%, which is expected since this is a binary classification task and does not require  precise AoA values.

\vskip 0.05in\noindent{\it Localization protocol round-trip time measurement.} {Finally,  we ran our distributed protocol  with different number of devices in a testbed  and measured the round-trip time for each  2D localization protocol run. For each  configuration with a specific number of devices, we ran the protocol 40 times and calculated the average round time for the protocol. For 3, 4, 5, 6 and 7 devices, the mean round times were  1.2, 1.6, 1.9, 2.2, and  2.5 seconds, respectively.}

\begin{figure}[t!]
    \includegraphics[width=.23\textwidth]{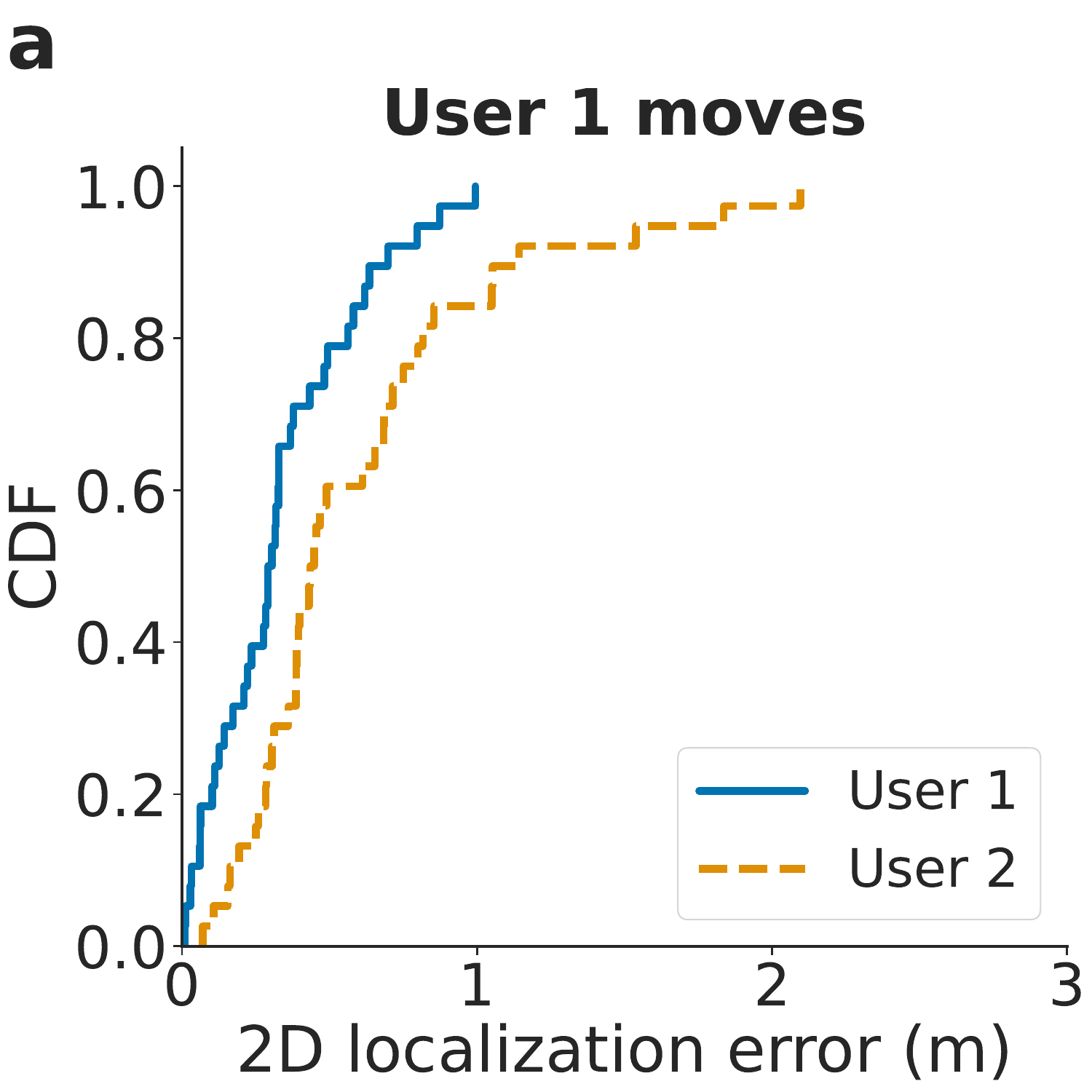}
    \includegraphics[width=.23\textwidth]{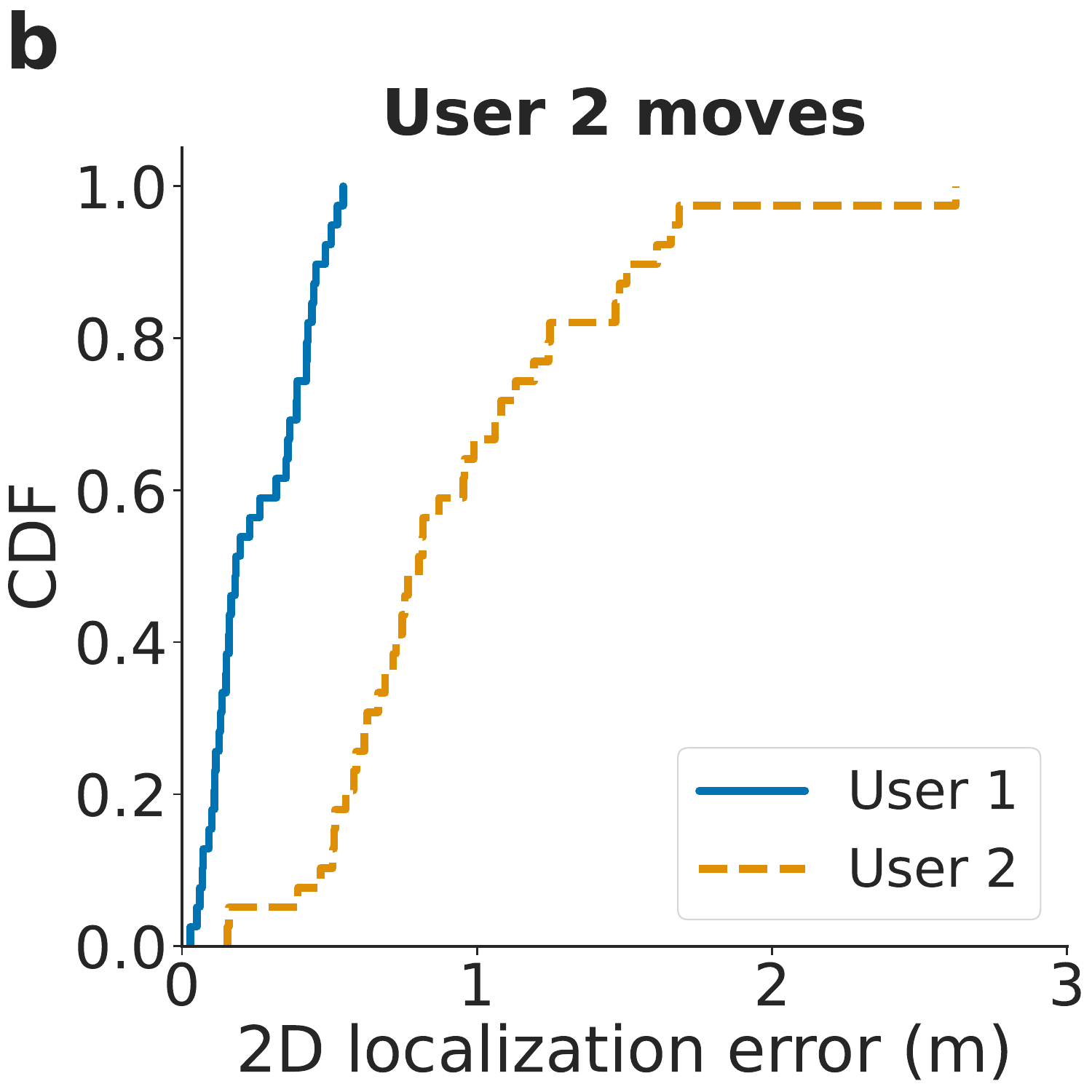}
    \vskip -0.15in
\caption{{2D localization errors for moving devices.}}
\vskip -0.2in
\label{fig:motion2d}
\end{figure}

\section{Related work}
 While there is prior  work on  underwater communication and messaging~\cite{sigcomm22,waterbackscatter}, here we focus on localization.

\vskip 0.05in\noindent{\bf Anchor-based underwater localization.} There has been prior interest in achieving underwater tracking for dive computers, sensors  and robots~\cite{tracking1, tracking2,tracking3,tracking4,phonerange, tracking5,tracking6,tracking7,tracking8,sensor1,vickery1998acoustic, cario2021accurate, ullah2019efficient,morework}. These proposals use time of arrival~\cite{single1}, time difference of arrival~\cite{reverse1}, angle of arrival~\cite{aoa} or signal strength~\cite{rssi} to estimate distances and angles from known anchor buoys (see~\cite{survey} for a detailed survey). 

 \cite{diver1} uses  hydrophone devices on the surface as beacons with known locations. The diver then uses a hand-held display connected to an acoustic communication module~\cite{gpsdiver1}.  \cite{ekf1} proposes the use of an underwater pinger with  a very high precision clock that is synchronized to GPS, prior to deployment.  The surface buoys may be equipped with GPS that can be translated to underwater GPS~\cite{gps1}.   \cite{single1}   computes the position based on time of arrival measurements for an underwater autonomous vehicle. \cite{directional1} proposes the use of directional beacons for localization. \cite{mobicom22} deploys a large number of visible passive tags which are used as anchors  for underwater localization; in addition to requiring a large number of tags, visual systems have a smaller underwater range than acoustic systems.

 Multi-hop localization~\cite{dvhop,centroid,multihop,multihop1} has been explored for  terrestrial and underwater ad-hoc sensor networks. Here, the distance to each anchor device is computed  using intermediary devices at different hops.  While these works are  theoretical in nature, they provide avenues for localizing devices in ad-hoc  networks. These methods however are primarily designed for anchor-based sensor networks, which is in contrast to our anchor-free design. 

\vskip 0.05in\noindent{\bf Anchor-free underwater localization.} 
\tuochao{This problem formulation is  under-explored for underwater settings.  \cite{anchorfree2} investigates ``active-restricted''  sensors that are anchored to the bottom of the sea and leverages the limited motion within a hemispherical area centered at the anchor. \cite{anchorfree_3} combines a discovery protocol and pairwise ranging to localize the relative coordination of the underwater sensor networks. \cite{anchorfree_1} propose an algorithm to account for the motion effect of acoustic nodes and improve the self-localization accuracy. Our work is more focused on some practical problems:  outlier detection, missing links, rotational and flipping ambiguities.} 

Anchor-free localization has also been explored for autonomous underwater vehicles to correct for their accumulated inertial errors~\cite{auvfree1,auvfree2,auvfree3}.Autonomous underwater vehicles however can more precisely control and know their motion, velocity and instantaneous direction, which is difficult to do using a smart device that is wore on a mobile human. \tuochao{For example, \cite{anchorfree_2} leverage the high-precision clock (drift rate < 1 ms over 14 hrs) and Doppler sonar to track the robot positions, which are not available on mobile devices. }
Finally some commercial products~\cite{ariadna} claim to achieve anchor-free localization using custom hardware with precise inertial sensors. Our evaluation with smart device IMUs confirmed prior observations that they  drift within a few seconds~\cite{lili}, making them challenging to use for anchor-free  localization.

\vskip 0.05in\noindent{\bf Underwater Synchronization.} 
\tuochao{ For  time of arrival  localization,  clock synchronization is critical. \cite{anchorfree_2} avoids  clock drifting by deploying high-precision clocks (drift rate < 1 ms over 14 hrs). \cite{clock2}  investigates the use of expensive atomic clocks for  underwater localization.  \cite{mao2016cat, millisonic} calibrate the clock drifting during the system initialization. However, such calibration requires known initial positions of devices, which introduces initialization overhead. Moreover, the residual estimation error will still lead to drifting with time. \cite{sync1, sync2, sync3} apply a two-way timestamp exchanging protocol for  synchronized among different nodes. \cite{sync4} utilizes the Doppler shift to estimate the clock skew and assumes the estimated velocity keeps constant within the sync interval. These methods require frequent message exchange to estimate the time offset and assume a higher bandwidth than is available on smart devices. In our system, instead of frequently synchronizing among devices with audible queries,   synchronization only happens when the leader initiates localization, which is more appropriate for our  application. }


\vskip 0.05in\noindent{\bf In-air localization.}   \cite{peng2007beepbeep,zhang2012swordfight} achieve in-air 1D acoustic ranging  between two phones.  2D acoustic localization uses anchor devices as beacons~\cite{cricket} and  microphone and speaker arrays~\cite{zhang2017soundtrak,millisonic,mao2016cat} to perform either  triangulation at distances of a few meters or AoA estimation.  In contrast, our target application requires localization at distances of 30-40~m. Accurate underwater AoA estimation requires microphone separation on the order of a meter at 1-4~kHz, which is an order of magnitude larger than a mobile device.  

Distributed  localization has been theoretically explored for ad-hoc anchor-free sensor networks~\cite{hari,anchorfree1,capkun}. \cite{hari,capkun} use network embeddings but do not address rotational and flipping ambiguities, while \cite{anchorfree1} assumes that each device is capable of measuring the angle of arrival from other devices.  The closest to our work  are   \cite{erdelyi2018sonoloc,teller} which achieve distributed in-air acoustic localization. \cite{teller} is designed for a  network with 16-40 sensors but assumes that the pair-wise distance errors are 1-5~cm, which is an order of magnitude lower than in underwater scenarios.  \cite{erdelyi2018sonoloc}  is limited in three key aspects: 1) it requires at least 10-15 devices to provide localization results. A dive party is typically much less than 10   divers,  2) it does not address rotational and flipping ambiguities, and 3) it explicitly assumes that all devices are in range of each other and  no occlusions exist between any device pairs.


\section{Discussion and Conclusion}

We present the first underwater acoustic positioning system for smart devices. Our software system achieves 3D positioning on commodity devices  without  external infrastructure. Here, we discuss the limitations of our current design.


\vskip 0.05in\noindent{\it When does it fail?} Sometimes our localization system, like any wireless scheme, produces high localization errors, as demonstrated by the long tail in Fig.~\ref{fig:block_drop}. A topology-based design must fulfill three key requirements: the network must be connected, each node must have at least three links, and the links must be "well distributed" across the nodes (see \xref{sec:localization}). Additionally, if a large number of pairwise distances  are erroneous, or if we encounter degenerate cases where say all devices are in a straight line, the localization results may be impacted. Finally, our approach necessitates at least three divers. With an increase in the number of divers, the design becomes more resilient to erroneous pairwise  measurements. With  two divers, we can only provide ranging information.

\vskip 0.05in\noindent{\it Two-hop communication.} Our distributed timestamp protocol (\xref{sec:protocol}) is designed to function even when not all devices are within range of the leader device. Additionally, our missing link evaluation (Fig.~\ref{fig:block_drop}) drops the direct link to the leader. However, our current implementation  assumes that all devices are only one hop away from the leader device, even if the link quality is imperfect. 

\vskip 0.05in\noindent{\it Localization  versus tracking.}  Our  system enables a device to calculate the 3D locations of other devices in relation to itself. However, this is initiated by the user and is not a continuous tracking system. This is a deliberate design choice, as it minimizes the duration of acoustic signals  underwater. Future work is necessary to develop a continuous tracking system that could potentially perform sensor fusion with other sensors, without continuous use of acoustics.


\vskip 0.05in\noindent{\it Audibility.} 
As  with prior work~\cite{sigcomm22}, we use 1-5~kHz  for acoustic pairwise distance estimation.  These are in the  human
hearing range, similar to many of the commercial and research modems~\cite{waterbackscatter,lowfreq-modem,lowfreq-modem1}.   To limit the time duration  we use acoustic signals underwater, our  system is designed to be an user-initiated action that is only performed when the leader wants to know the positions of their dive group and is not designed for continuous tracking. 

\vskip 0.05in\noindent\textcolor{red}{{\it Visual odometry.} A  design decision we made in our paper is to not use cameras for localization despite they being ubiquitous on smartphones.  We opted not to use cameras since our goal is to create a design that  works for wearables like smart watches  which are more likely than phones to be  used in underwater scenarios. Cameras unfortunately are not yet common on smart watches. Light-based methods are also susceptible to turbid water~\cite{lacovara2008high}. 
}

\vskip 0.05in\noindent\textcolor{red}{{\it Diver evaluations.} 
In this paper, we mainly use ropes and long sticks to put the devices underwater during evaluation. Further work is required to evaluate our system with  divers in real-world dives (seaside or deep ocean) and with multiple divers constantly moving.
}


\vskip 0.05in\noindent{\bf Acknowledgments.}
The researchers are funded by  the  Moore Inventor Fellow award \#10617 and the National Science Foundation. We thank our shepherd, Fadel Adib,  Malek Itani, Antonio Glenn, Bandhav Veluri and Maruchi Kim for their feedback.

\vskip 0.05in\noindent This work does not raise  ethical issues.

\bibliographystyle{ACM-Reference-Format}
\bibliography{ms}

\section*{Appendix} 

\vskip 0.05in\textbf{Low-level audio timing.} 
Our goal is to ensure that the replying device can send a preamble at a precise sample index in the future that  corresponds to  $t_{reply}$, after the signal from the sender arrives at the device. To map the sample indices to time, we need to look into how Android transmits and records sound at a low level. The low-level OpenSL ES audio library in Android  exposes access to the speaker and microphone audio sample buffers. Specifically, the library provides the ability to directly write audio samples to a future speaker buffer even during speaker playback. During runtime, the library executes a \textit{CallBack} function when the microphone buffer is full or the speaker buffer is empty. In this way, we can acquire a continuous data stream for microphone data and another continuous data stream for speaker data.  Thus, the sample indices in the microphone  and speaker streams have a linear relationship with   timestamps:
\begin{equation}
\begin{aligned}
& t_s(n) = n/f_s^s + t_s^0 ,   
& t_m(m) = m/f_s^m + t_m^0 
\end{aligned}
\label{eq:linear}
\end{equation}
Here $m$ and $n$ are the sample indices in the microphone and speaker data streams. $t_s(n)$ is the timestamp when  sample $n$ in the  buffer is send out by the speaker and $t_m(m)$ is the timestamp when  sample $m$ arrives in the microphone  buffer. $t_s^0$ is the initial timestamp when the first sample in the stream is sent by the speaker, and $t_m^0$ is the initial timestamp when the first sample in the stream arrives at the microphone. $f_s^s$ and $f_s^m$ are the  sampling rates for the speaker and microphone, which may not be exactly  our desired sampling rate ($f_s$ = 44.1~kHz). We assume that $f_s^s = f_s/(1-\alpha)$ and $f_s^m = f_s/(1-\beta)$, where $|\alpha| \ll 1$ and $|\beta| \ll 1$.

\vskip 0.05in\noindent\textbf{Self-synchronizing speaker and microphone streams.}
As shown in Fig.~\ref{fig:pipeline2}, we do not know the exact timestamp $t_3$ when the preamble arrived at the microphone of device B. Instead, we only know the sample index $m_2$ of the recorded preamble in the microphone stream. At the speaker side, we also cannot directly know the exact timestamp $t_4$ when the speaker sends the signal, but we can control the sample index $n_2$  in the speaker stream. We define the $\delta_2$ is the propagation delay from the speaker to its own microphone. { We define $t_{reply}$ as the time interval between the arrival of signal from phone A and the arrival of its own  signal at the phone B's microphone. According to Fig.~\ref{fig:pipeline2} we have $t_{reply} = t_5 - t_3 = t_4 + \delta_2 - t_3$. Combining this with  Eq.~\ref{eq:linear}, we have}
\begin{equation}
\begin{aligned}
t_{reply} & = t_4 + \delta_2 - t_3= n_2/f_s^s + t_s^0 + \delta_2 - m_2/f_s^m - t_m^0 
\end{aligned}
\label{eq:reply}
\end{equation}
 The microphone  and speaker buffers work  separately, and there is no guarantee of the relative order between the two buffers. In other words, the initial offsets  $t_s^0$ and $t_m^0$ can be different each time we open the streams. To address this,  once we open the microphone and speaker data streams, we do not close them so as to  keep the offset, $t_s^0-t_m^0$, constant. We write zeros to the speaker stream when we are transmitting nothing to keep the buffer full. Further, after initializing the two streams, the speaker sends a calibration signal (green in Fig.~\ref{fig:pipeline2}) to estimate this offset. We write the calibration signal to the sample index $n_1$ in the speaker stream. Then the microphone stream would receive this calibration signal at sample index $m_1$. 
{The propagation time of the  calibration signal from the speaker to the microphone on device B is $(t_2-t_1)$ (i.e., $\delta_2$). By applying Eq.~\ref{eq:linear}, we get:}
\begin{equation}
\begin{aligned}
& t_2-t_1 = m_1/f_s^m + t_m^0 - n_1/f_s^s - t_s^0 = \delta_2 
\end{aligned}
\label{eq:delta}
\end{equation}
\begin{figure}[t!]
    \includegraphics[width=.47\textwidth]{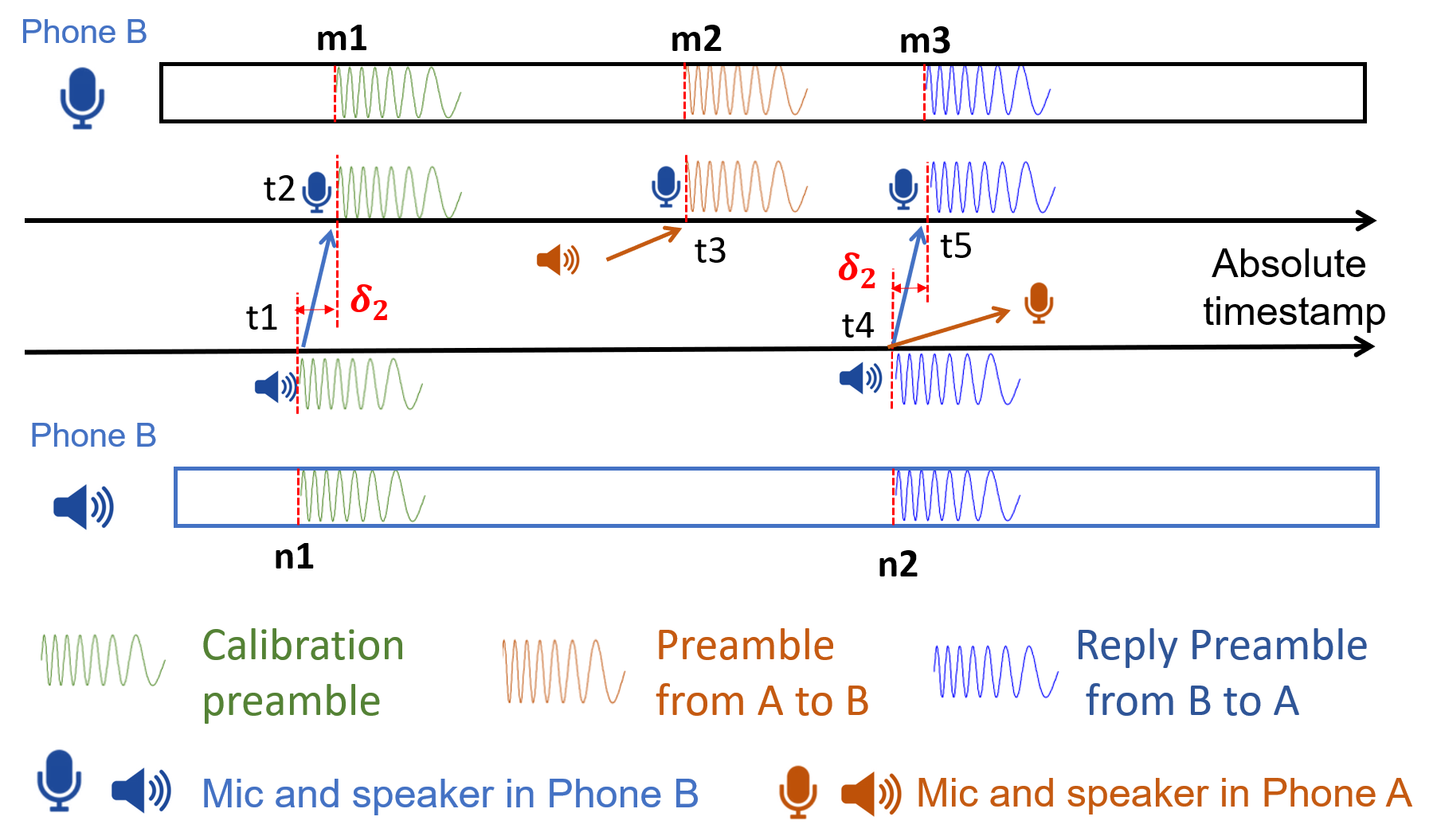}
    \vskip -0.15in
\caption{{\bf Mapping  buffer samples to absolute time.}}
\vskip -0.15in
\label{fig:pipeline2}
\end{figure}
{Now, we compute  the  offset $(n1 - m1)$ between the  microphone and speaker after initial calibration, which can be used for $(t_s^0-t_m^0)$  compensation. }
After initial calibration, our goal is when device B detects the start of the signal from device A at the sample index $m_2$ in Fig.~\ref{fig:pipeline2},  device B will write the reply signal at the sample index $n_2$ in the speaker stream, such that they are separated in time by the desired gap, $t_{reply}^0$, after adjusting for buffer delays. So, by compensating the indices offset acquired from calibration, we set $n_2$ to, 
\begin{equation}
\begin{aligned}
n_2 = m_2 + (n_1 - m_1) + fs\cdot t_{reply}^0, 
\end{aligned}
\label{eq:n2}
\end{equation}
Here $f_s$ is the desired sampling rate. However the real reply interval $t_{reply}$ is shown in Eq.~\ref{eq:reply}. The difference between the real and desired reply times is,
\begin{equation}
\begin{aligned}
& t_{reply} - t_{reply}^0 = n_2/f_s^s + t_s^0 - m_2/f_s^m - t_m^0 - t_{reply}^0 + \delta_2\\
\label{eq:diff}
\end{aligned}
\end{equation}
{By combining Eqs.~\ref{eq:delta} and~\ref{eq:diff} and using the  relationship between $f_s$, $f_s^m$, $f_s^s$, we can rewrite the above equation as,}
\begin{equation}
\begin{aligned}
& t_{reply} - t_{reply}^0 
& = - \alpha t_{reply}^0 + \frac{(m_2-m_1)(\beta - \alpha)}{f_s}
\end{aligned}
\end{equation}

The key observation here is that the main error source is from the difference between the actual sampling rate and the nominal sampling rate.  {$\alpha$ for  Android devices  is around 1-80 ppm~\cite{guggenberger2015analysis}.  As for the second term, while  $\beta-\alpha$ is the clock drifting difference between speaker and microphone clock, which is usually quite small in most Android phone. To avoid the second error term accumulate as $m_2 - m_1$ increases, we can utilize the response signal of this device to re-calibrate the offset between the speaker and microphone streams.}

\vskip 0.05in\noindent{\it SNR measurement.} \tuochao{We measure the Signal-to-Noise Ratio (SNR) at 10, 20, and 28~m at the boathouse. During the measurement, there were  people fishing and boating nearby. To estimate the SNR, we send a preamble consisting of 8 OFDM symbols from 1-5 KHz. We compute the SNR for each subcarrier by applying frequency-domain MMSE channel estimation ~\cite{sigcomm22}. Fig.~\ref{fig:main_result1} shows the estimated SNR for each subcarrier between 2 Samsung S9 phones. }

\begin{figure}[t!]
    \includegraphics[width=.3\textwidth]{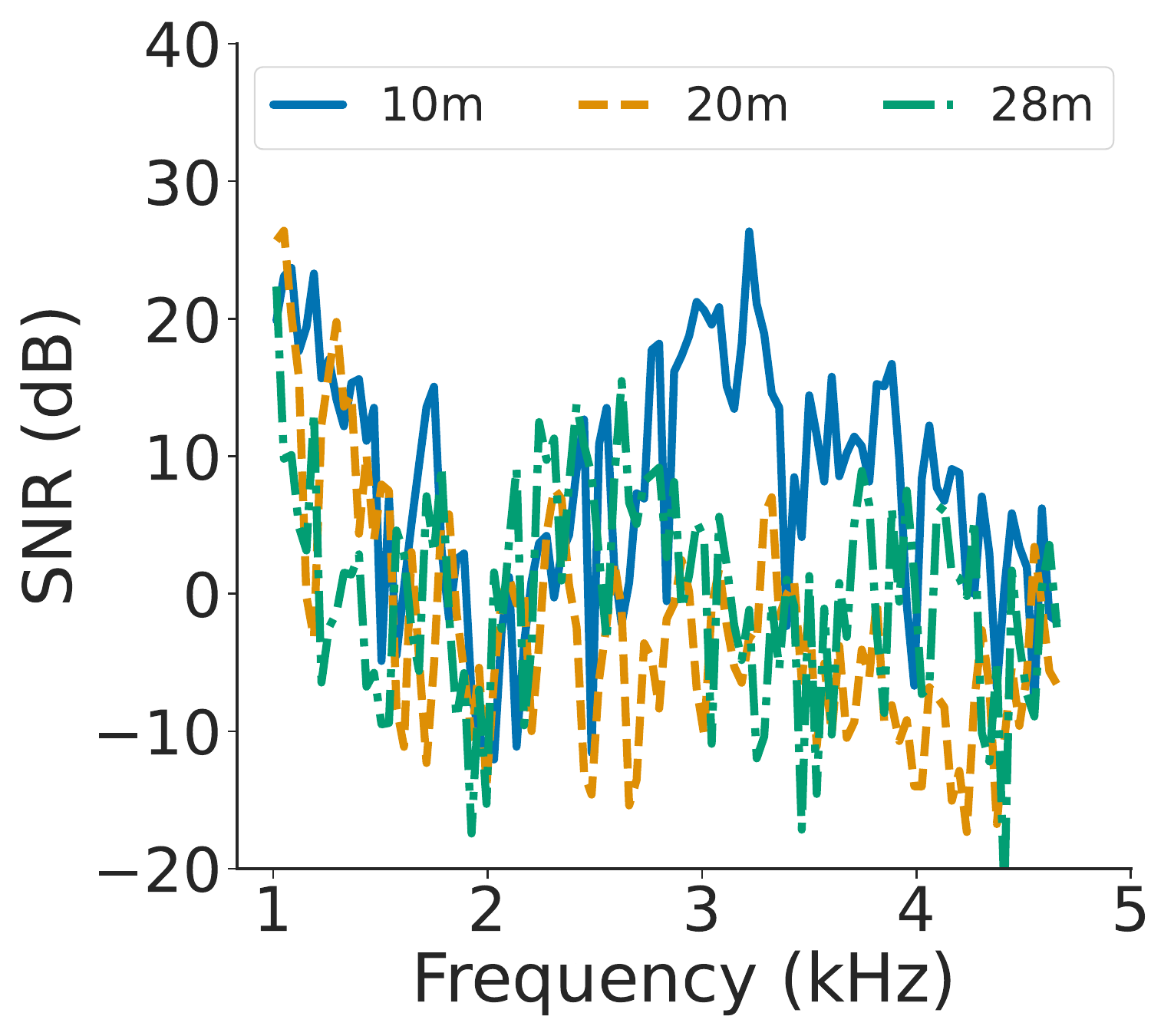}
   \vskip -0.2in
\caption{SNR estimation between two Samsung S9 devices at different  distances. }
\vskip -0.15in
\label{fig:main_result1}
\end{figure}

\end{document}